\documentclass[onecolumn,preprint,showpacs,superscriptaddress,aps]{revtex4-1}

\usepackage[utf8]{inputenc}
\usepackage{mathtools}
\usepackage{graphicx}
\usepackage{caption}
\usepackage{subcaption}
\usepackage{xcolor}
\usepackage{soul}
\usepackage{float}
\usepackage{amsmath}
\usepackage{amssymb}
\usepackage{amsfonts}
\usepackage{appendix}
\usepackage{orcidlink}

\usepackage{dcolumn}% Align table columns on decimal point
\usepackage{bm}% bold math
\usepackage{epstopdf}
\usepackage{relsize}
\usepackage{placeins}
\usepackage{array}
\usepackage{comment}
\usepackage{hyperref}
\usepackage{booktabs}

\usepackage{cleveref}

\crefname{equation}{Eq.}{Eqs.}
\crefname{figure}{Fig.}{Figs.}
\crefname{table}{Tab.}{Tabs.}

\begin{document}

\title{Charged scalar boson in Melvin universe}

\author{L. G. Barbosa \orcidlink{0009-0007-3468-3718}}
\email{leonardo.barbosa@posgrad.ufsc.br}
\affiliation{Department of Physics - Federal University of \\ Santa Catarina; 88.040-900, Florianópolis, SC, Brazil}

\author{L. C. N. Santos \orcidlink{0000-0002-6129-1820}}
\email{luis.santos@ufsc.br}
\affiliation{Department of Physics - Federal University of \\ Santa Catarina; 88.040-900, Florianópolis, SC, Brazil}

\author{J. V. Zamperlini \orcidlink{0009-0002-9702-1555}}
\email{joao.zamperlini@posgrad.ufsc.br}
\affiliation{Department of Physics - Federal University of \\ Santa Catarina; 88.040-900, Florianópolis, SC, Brazil}

\author{F. M da Silva  \orcidlink{0000-0003-2568-2901}}
\email{franciele.m.s@ufsc.br}
\affiliation{Department of Physics - Federal University of \\ Santa Catarina; 88.040-900, Florianópolis, SC, Brazil}

\author{C. C. Barros Jr. \orcidlink{0000-0003-2662-1844}}
\email{barros.celso@ufsc.br}
\affiliation{Department of Physics - Federal University of \\ Santa Catarina; 88.040-900, Florianópolis, SC, Brazil}

\begin{abstract}
\textcolor{black}{This work investigates the dynamics of a charged scalar boson in the Melvin universe by solving the Klein-Gordon equation with minimal coupling in both inertial and non-inertial frames. Non-inertial effects are introduced through a rotating reference frame, resulting in a modified spacetime geometry and the appearance of a critical radius that limits the radial domain of the field. Analytical solutions are obtained under appropriate approximations, and the corresponding energy spectra are derived. The results indicate that both the magnetic field and non-inertial effects modify the energy levels, with additional contributions depending on the coupling between the rotation parameter and the quantum numbers. A numerical analysis is also presented, illustrating the behavior of the solutions for two characteristic magnetic field scales: one that may be considered extreme, of the order of the ones proposed to be produced in heavy-ion collisions and another near the Planck scale.}
\end{abstract}

\maketitle

\section{Introduction}
In recent years, several physical systems in nature with strong magnetic fields have been investigated. Some of these systems include magnetars, which are neutron stars with very strong magnetic fields up to $10^{15}-10^{16}$G \cite{thompson1995soft,mereghetti2015magnetars,Lopes:2014vva}, ultra-relativistic heavy-ion collisions where it is expected that these fields reach the order of $10^{19}\: $G \cite{bzdak2012event,voronyuk2011electromagnetic,gursoy2014magnetohydrodynamics} and, in general, quantum electrodynamics (QED) and quantum chromodynamics (QCD) systems where phenomena driven by the influence of magnetic fields is significant \cite{adhikari2024strongly}.

Another relevant type of investigation in this field is to study models of classical and quantum dynamics of particles that are under the effect of such intense magnetic fields. In the case of scalar particles, the Klein-Gordon equation is widely used to describe the dynamics of a particle or quantum field. In this context, it is well known in the literature that the curvature of spacetime makes it difficult to solve the resulting differential equations due to the intricate form it takes in curved geometry. In this regard, the spacetime of a cosmic string is an example of a spacetime with a high degree of symmetry that allows the existence of exact solutions and has been explored in a number of works. For example, the problem of spin-0 and spin-1/2 particles under the effect of a homogeneous magnetic field was studied in \cite{figueiredo2012relativistic}, where the authors considered the cosmic string geometry and scalar potentials. The Aharonov-Bohm potential is another type of potential that is commonly studied in wave equations in curved spacetimes. This effect demonstrates that charged particles can be influenced by electromagnetic potentials even in regions where the electromagnetic field is zero \cite{cavalcanti,boumali}. 

In addition to scalar fields, other systems studied in the cosmic string spacetime include non-relativistic \cite{wang2015exact,muniz2014landau,ikot2016solutions,ahmed2023effects} and fermionic systems \cite{inercial10,inercial8,string8,string7,string1,hosseinpour2015scattering,marques2005exact,bakke2018dirac,hosseinpour2017scattering,cunha2020dirac,lima20192d}. The study of particle dynamics in geometries originating from the solutions of Einstein's equation in the presence of magnetic fields has gained attention in recent years. The Dirac equation in the presence of a curved geometry due to a constant magnetic field was solved first in \cite{santos}, where the authors study the dynamics of fermions in the Melvin universe and show that the energy spectrum undergoes a shift in a high-energy environment. The dynamics of test charged matter waves in this static geometry have been studied recently in \cite{bini2022static}. In particular, the occurrence of cosmic double-jet configurations in this universe has been discussed. 

\textcolor{black}{Generalizations of Melvin-type solutions that incorporate a cosmological constant $\Lambda$ have been proposed to study the combined influence of magnetic fields and vacuum energy on the structure of spacetime. Such extensions, often referred to as Bonnor-Melvin-$\Lambda$ spacetimes, have been considered in the context of cylindrically symmetric configurations supported by electromagnetic fields and a nonzero $\Lambda$ \cite{Zofka:2019yfa}. More recently, scalar fields and relativistic oscillators have been investigated in these geometries, revealing that the cosmological constant can significantly modify the energy spectrum and affect the behavior of the wave function when additional scalar and vector potentials are present \cite{Barbosa:2023gxl, Barbosa:2023rmq}.}

Solutions of wave equations in spacetimes with geometry influenced by magnetic fields can be related to well-known solutions in flat spacetime. For instance, the solution studied in \cite{santos} provides additional terms due to the curvature of spacetime. Such terms can be neglected in the limit where the geometry becomes flat and the final result agrees with the usual Landau levels \cite{melrose1983quantum}. Taking this into account, we intend to expand the study of quantum particle dynamics in the Melvin universe and consider scalar particles. Furthermore, we will obtain a new configuration of the Melvin metric associated with a non-inertial reference frame. In this work, non-inertial effects are introduced into the system through the standard procedure in the angular sector $\phi = \chi + \omega t$ that has provided interesting results for the systems studied in the literature \cite{bakke2012noninertial,mota2014noninertial,ahmed2023non,santos2023some,santos2017scalar,santos2018relativistic,santos2023non}, allowing in many cases exact solutions.
%aqui citar santos2018relativistic

The structure of this paper is as follows: in Section (\ref{Melvin_Metric}), we revisit the Melvin metric as a solution of the Einstein-Maxwell equations. In Section (\ref{Non_Inertial}), we introduce a non-inertial reference frame through a coordinate transformation. We derive the modified line element and obtain the corresponding critical radius that bounds the physical region of the spacetime. In Section (\ref{Klein_Gordon_Melvin}), we analyze the Klein-Gordon equation in the Melvin universe without non-inertial effects and derive the energy spectrum for a charged scalar boson. In Section (\ref{Klein_Gordon_Non_Inertial_Melvin}), we incorporate non-inertial effects and obtain the modified radial equation and its solutions. We explore two regimes: one in which the radial domain extends to infinity, and another in which the critical radius is finite and a hard-wall boundary condition must be imposed. In Section (\ref{Conclusions}), we summarize our findings and discuss the influence of the magnetic field and non-inertial effects on the energy spectrum.

{\color{black} Throughout the paper, we employ the use of natural units, where $\hbar=c=1$ and $G=m_P^{-2}=l_P^2$, where $m_P$ and $l_P$ are the Planck mass and Planck length, respectively. Additionally, the elementary charge becomes $q_e=\sqrt{4\pi\alpha}$, where $\alpha\approx1/137$, is the fine structure constant.}

\section{Melvin Metric}\label{Melvin_Metric}

In this section, we revisit the Melvin metric {\color{black}\cite{bonnor1954PPSA,melvin1964PhL,melvin1965PhRv,thorne1965PhRv}, which describes the curved spacetime of a static magnetic field in General Relativity (GR). To encounter the Melvin metric, we need to solve the Einstein-Maxwell equations.} We begin with Einstein's {\color{black} field} equation 
\begin{equation}\label{EqEinstein}
G_{\mu \nu}=R_{\mu\nu} - \frac{1}{2} R g_{\mu\nu} = 8\pi G T_{\mu\nu},
\end{equation}
{\color{black} where the constant magnetic field will be accounted by $T_{\mu \nu}$,} the electromagnetic energy-momentum tensor 
\begin{equation}
T_{\mu\nu}=\frac{1}{4\pi}\left(F_{\mu}^{\;\;\alpha}F_{\nu\alpha}-\frac{1}{4}F_{\alpha\beta}F^{\alpha\beta}g_{\mu\nu}\right),
\end{equation}
where the Maxwell tensor \(F_{\mu \nu}\) is defined in terms of the electromagnetic potential \(A_{\mu}\) and the covariant derivative \(\nabla_{\mu}\)
\begin{equation}
    F_{\mu\nu} = \nabla_{\nu} A_{\mu} - \nabla_{\mu} A_{\nu},
\end{equation}
and satisfies Maxwell's equation in curved spacetime
\begin{equation}\label{MaxwellI}
\nabla_{\mu} F^{\mu\nu} = 0.
\end{equation}

{\color{black}In this work, we are interested in a spacetime} with cylindrical symmetry, {\color{black} which can be described by} the following line element:
\begin{equation}\label{metric}
   ds^{2}=H\left(r\right)\left(-dt^{2}+dr^{2}+dz^{2}\right)+\frac{r^{2}}{H\left(r\right)}d\phi^{2}, 
\end{equation}
where \(t, z \in \mathbb{R}\), \(r \in \mathbb{R}_{+}\), and \(\phi \in [0, 2\pi)\). The magnetic field is assumed to be aligned with the symmetry axis, and generated by an electromagnetic potential {\color{black} as follows}
\begin{equation}
  A_{\mu}=A_{\phi}\left(r\right)\delta_{\mu}^{\phi}, \label{potentialA}
\end{equation}
such that{\color{black}, the Maxwell tensor will be given by}
\begin{equation}
 F_{r\phi}=\partial_{r}A_{\phi}\left(r\right)=B\left(r\right).
\end{equation}

%Considering the line element and
{\color{black}Applying the metric tensor to \cref{MaxwellI}, we obtain} Maxwell's equations written in terms of the Laplace–Beltrami operator,
\begin{equation}
    \frac{1}{\sqrt{-g}}\partial_{\nu}\left(\sqrt{-g}g^{\alpha\mu}g^{\beta\nu}F_{\alpha\beta}\right)=0.
\end{equation}
{\color{black} Replacing the line element of \cref{metric} and the Maxwell tensor of \cref{potentialA} in the equation above, we obtain that} the magnetic field \(B\left(r\right)\) can be expressed in terms of the metric components as
\begin{equation} \label{eqB}
    B\left(r\right)=\partial_{r}A_{\phi}=\frac{B_{0}r}{H\left(r\right)},
\end{equation}
where \(B_0\) is an integration constant that may be interpreted as the magnetic field strength. {\color{black}To find the solution to the Einstein field equations,} next, we consider the components of {\color{black}the Einstein tensor} \(G_{\mu \nu}\) {\color{black}that do not vanish}:
\begin{equation}
   G_{tt}=G_{rr}=-G_{zz}=-\frac{H^{'}\left(r\right)}{4rH^{2}\left(r\right)}\left(rH^{'}\left(r\right)-4H(r)\right),
\end{equation}
\begin{equation}
    G_{\phi\phi}=\frac{r^{2}}{4H^{4}(r)}\left(4H(r)H^{''}\left(r\right)-3\left(H^{'}\left(r\right)\right)^{2}\right),
\end{equation}
where the prime denotes differentiation with respect to the radial coordinate. {\color{black}We also consider the} nonzero components of the energy-momentum tensor, which are given by
\begin{equation}
  8\pi GT_{tt}=8\pi GT_{rr}=-8\pi GT_{zz}=\frac{GH\left(r\right)}{r^{2}}\left(A_{\phi}^{'}\left(r\right)\right)^{2},
\end{equation}
\begin{equation}
   8\pi GT_{\phi\phi}=\frac{G}{H\left(r\right)}\left(A_{\phi}^{'}\left(r\right)\right)^{2}.
\end{equation}

{\color{black}Then,} equating the corresponding components and considering the relation that characterizes the magnetic field {\color{black}given by \cref{eqB}}, we obtain {\color{black} the following equations:}
\begin{equation} \label{field1}
    \left(H^{'}\left(r\right)\right)^{2}-\frac{4H(r)}{r}H^{'}\left(r\right)+4GB_{0}^{2}H(r)=0,
\end{equation}
\begin{equation}\label{field2}
    4H(r)H^{''}\left(r\right)-3\left(H^{'}\left(r\right)\right)^{2}-4GB_{0}^{2}H(r)=0.
\end{equation}

Substituting {\color{black} \cref{field1} into \cref{field2}, or vice versa,} yields  {\color{black}the following second-order differential equation}
\begin{equation}
    H^{''}\left(r\right)-\frac{3}{r}H^{'}\left(r\right)+2GB_{0}^{2}=0,
\end{equation}
whose solution is given by
\begin{equation}
    H\left(r\right)=\left(1+\frac{1}{4}GB_{0}^{2}r^{2}\right)^{2}.
\end{equation}

{\color{black}Now, that we have obtained} the solution for \(H\left(r\right)\), the electromagnetic potential can be determined explicitly by direct integration {\color{black}of \cref{eqB}}:
\begin{equation}
    A_{\phi}=-\frac{2}{GB_{0}\sqrt{H\left(r\right)}}.
\end{equation}

To recover Melvin's solution {\color{black} in cylindrical coordinates \cite{santos}}, we redefine \(H\left(r\right)=\Lambda^{2}\left(r\right)\), so that the line element and the electromagnetic potential become
\begin{equation} \label{melvinfinal}
    ds^{2}=\Lambda^{2}\left(r\right)\left(-dt^{2}+dr^{2}+dz^{2}\right)+\frac{r^{2}}{\Lambda^{2}\left(r\right)}d\phi^{2},
\end{equation}
\begin{equation}\label{Lambdafinal}
    \Lambda\left(r\right)=1+\frac{1}{4}GB_{0}^{2}r^{2},
\end{equation}
\begin{equation} \label{afinal}
    A_{\phi}\left(r\right)=-\frac{2}{GB_0\Lambda\left(r\right)},
\end{equation}
which corresponds to the Melvin solution characterizing the geometry generated and maintained by an axially symmetric magnetic field {\color{black}along the $z$ direction}. It should be noted that at the origin $(r=0)$ the electromagnetic potential assumes the value
\begin{equation}
  A_{\phi}\left(0\right)=-\frac{2}{GB_0}.
\end{equation}
In the limit of vanishing magnetic field, the electromagnetic potential diverges. In order to obtain a regular potential in this limit, one may consider a gauge transformation for the electromagnetic potential
\begin{equation}
    A_{\mu}\rightarrow A_{\mu}+\partial_{\mu}\chi,
\end{equation}
with \(\chi=\frac{2\phi}{GB_0}\). Applying this gauge transformation leads the electromagnetic potential {\color{black} which may be rewritten in the following form:}
\begin{equation}
   A_{\phi}\left(r\right)=\frac{B_{0}r^{2}}{2\Lambda\left(r\right)},
\end{equation}
which vanishes at the origin and remains free of divergences in the process of taking the limit of the magnetic field going to zero.

Given a solution of the Einstein-Maxwell equations, an important quantity that characterizes the structure of the spacetime singularities is the Kretschmann scalar, computed from the components of the Riemann tensor as detailed in Appendix (\ref{Riemann_Tensor_Components})
\begin{equation}
   K\left(t,r,\phi,z\right)=R_{\mu\nu\alpha\beta}R^{\mu\nu\alpha\beta}=\frac{3G^{4}B_{0}^{8}r^{4}}{4\Lambda^{8}\left(r\right)}-\frac{6G^{3}B_{0}^{6}r^{2}}{\Lambda^{8}\left(r\right)}+\frac{20G^{2}B_{0}^{4}}{\Lambda^{8}\left(r\right)}.
\end{equation}

\begin{figure}[H]
    \centering
    \includegraphics[width=0.7\linewidth]{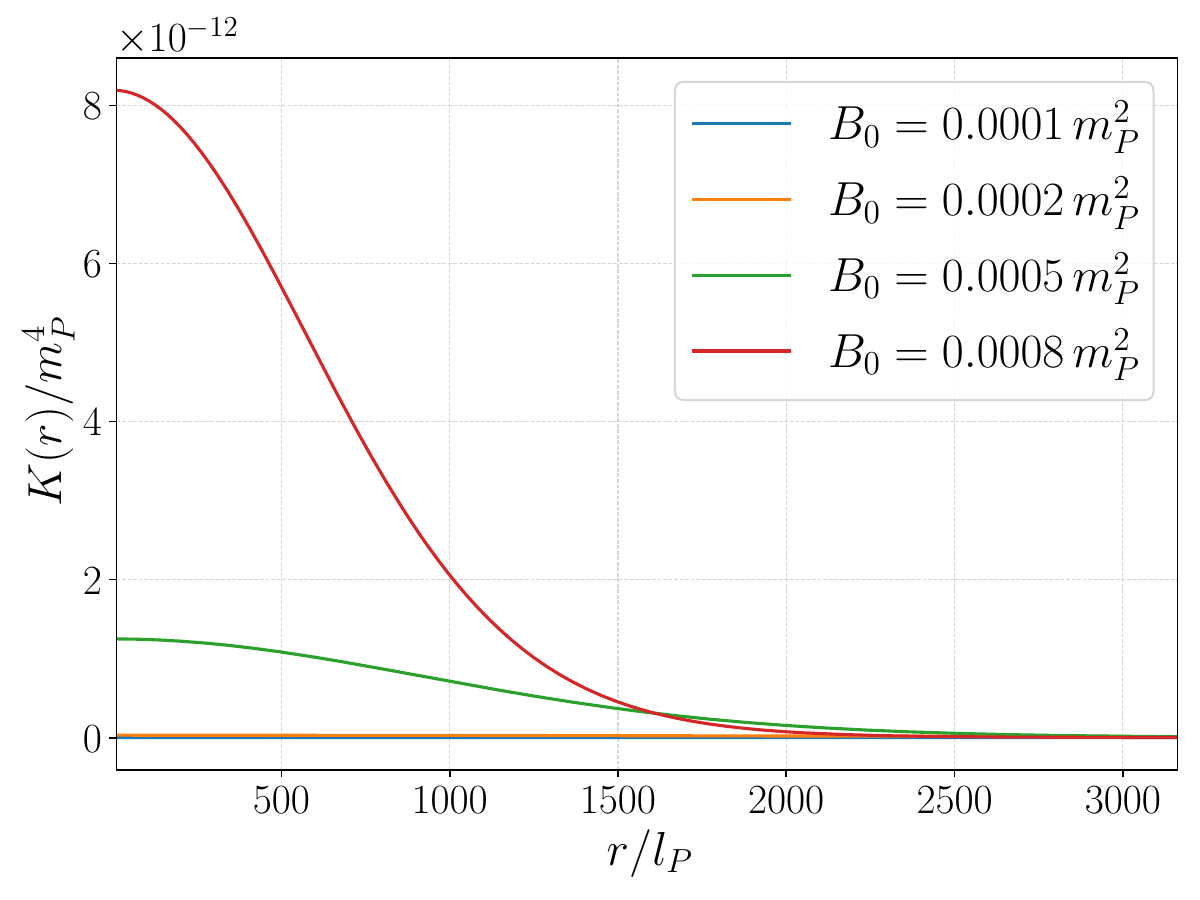}
    \caption{Kretschmann scalar against radial coordinate for different values of $B_0$.}
    \label{fig:kretschmann}
\end{figure}

In the limit of the radial coordinate approaching zero, the Kretschmann scalar tends to a finite value dependent on the magnetic field intensity, \(K\rightarrow20G^{2}B_0^{4}\). Conversely, as \(r\rightarrow\infty\), the Kretschmann scalar vanishes, \(K\rightarrow0\). 
{\color{black}In \autoref{fig:kretschmann} we can observe the behavior of the Kretschmann scalar $K$ near the origin ($r=0$) and for $r \rightarrow \infty$ for four different values of the magnetic field strength $B_0$. }

\section{Non-inertial Effects in the Melvin Universe}\label{Non_Inertial}

As it was pointed before, in this work
we are interested in studying non-inertial effects in the Melvin metric. To this end, we consider the following transformation in the angular sector of the solution:
\begin{equation} 
    \phi=\chi+\omega t,
\end{equation}
where \(\chi\) is the new angular variable and \(\omega\) is the angular velocity. Substituting {\color{black} the transformation above} into the line element {\color{black}of \cref{melvinfinal}}, we obtain
\begin{equation} \label{melvinfinalw}
    ds^{2}=-\left(\Lambda^{2}\left(r\right)-\frac{r^{2}\omega^{2}}{\Lambda^{2}\left(r\right)}\right)dt^{2}+\Lambda^{2}\left(r\right)dr^{2}+\frac{2\omega r^{2}}{\Lambda^{2}\left(r\right)}d\chi dt+\frac{r^{2}}{\Lambda^{2}\left(r\right)}d\chi^{2}+\Lambda^{2}\left(r\right)dz^{2},
\end{equation}
and applying the same transformation to the electromagnetic potential, we have
\begin{equation} \label{afinalw}
    A_{\mu}=A_{t}\left(r\right)\delta_{\mu}^{t}+A_{\chi}\left(r\right)\delta_{\mu}^{\chi}=\frac{\omega B_{0}r^{2}}{2\Lambda\left(r\right)}\delta_{\mu}^{t}+\frac{B_{0}r^{2}}{2\Lambda\left(r\right)}\delta_{\mu}^{\chi}.
\end{equation}

For a certain critical value $r_c$, the particle can have a speed greater than that of light, which is not allowed. Therefore, we will discuss how to identify this point and how to ensure that the solutions of the wave equation take this limit into account. Assuming that the signature of the line element remains unchanged, we impose the condition for a critical radius by setting \(g_{tt}=0\), which results in the equation
\begin{equation}\label{EqCriticalRadio}
\Lambda^{4}\left(r_{\text{c}}\right)=r_{\text{c}}^{2}\omega^{2}.
\end{equation}

\textcolor{black}{By retaining only the terms that are linear in the gravitational constant $G$, that is, up to the order $\mathcal{O}\left(G B_0^2 r^2\right)$, the function $\Lambda^N(r)$ can be approximated as 
\begin{equation}\label{AproxLambda}
\Lambda^N(r) \approx 1 + \frac{N}{4} G B_0^2 r^2.
\end{equation}
This approximation, \cref{AproxLambda}, is supported by the relevant physical parameters. In heavy-ion collisions, the magnetic field strength is expected to reach values of the order $B_0 \sim 10^{18}~\text{G}$, which corresponds to $B_0 \sim 10^{-2}~\text{GeV}^2$ in natural units. As a result, the dimensionless combination $G B_0^2 \sim 10^{-42}$ is extremely small. Consequently, higher-order contributions such as $\left(G B_0^2 r^2\right)^2$, $\left(G B_0^2 r^2\right)^3$, and so forth, can be considered negligible even for large values of $r$, which represent particles far from the origin, that is the region with significant values of the magnetic field which we are interested to investigate.} Solving \cref{EqCriticalRadio} under this approximation, the critical radius can be characterized in terms of the angular velocity and the magnetic field strength as
\begin{equation}\label{Criticalradio}
    r_{\text{c}}\approx\frac{1}{\sqrt{\omega^{2}-GB_{0}^{2}}}.
\end{equation}

Thus, in the description of phenomena in the Melvin spacetime with non-inertial effects, the radial coordinate must vary between zero and the critical radius, i.e., \(0<r<r_{\text{c}}\). 

\textcolor{black}{An analysis of the critical radius reveals that the minimum value for which the expression from \cref{Criticalradio} remains valid depends on the value of $B_0$ and is given by $\omega_{\text{min}} = \sqrt{G}B_{0}$.} \textcolor{black}{On the other hand, if we assume a sensible minimal length that $r_\mathrm{c}$ can reach, for instance $l_P$, then, $\omega$ will also be constrained by the magnetic field value, such that: $\omega_{\text{max}}=\sqrt{m_{\text{P}}^{2}+GB_{0}^{2}}$.} \textcolor{black}{This way, we obtain the following validity range for $\omega$:
\begin{equation}\label{OmegaInt}
\omega_{\text{max}}\geq\omega>\omega_{\text{min}}.
\end{equation}}

{\color{black} 
To visualize, we refer to \autoref{fig:combined-rcriticalvsOmega}, from which we can see the dependence of the critical radius on $\omega$, considering two examples, a magnetic field near the heavy-ion collisions scale on \autoref{fig:rcriticalvsomega-planckscalea}, and also near the Planck scale on \autoref{fig:rcriticalvsomega-planckscaleb}.

\begin{figure}[H]
    \centering
    \begin{subfigure}{0.49\textwidth}
        \centering
        \includegraphics[width=\linewidth]{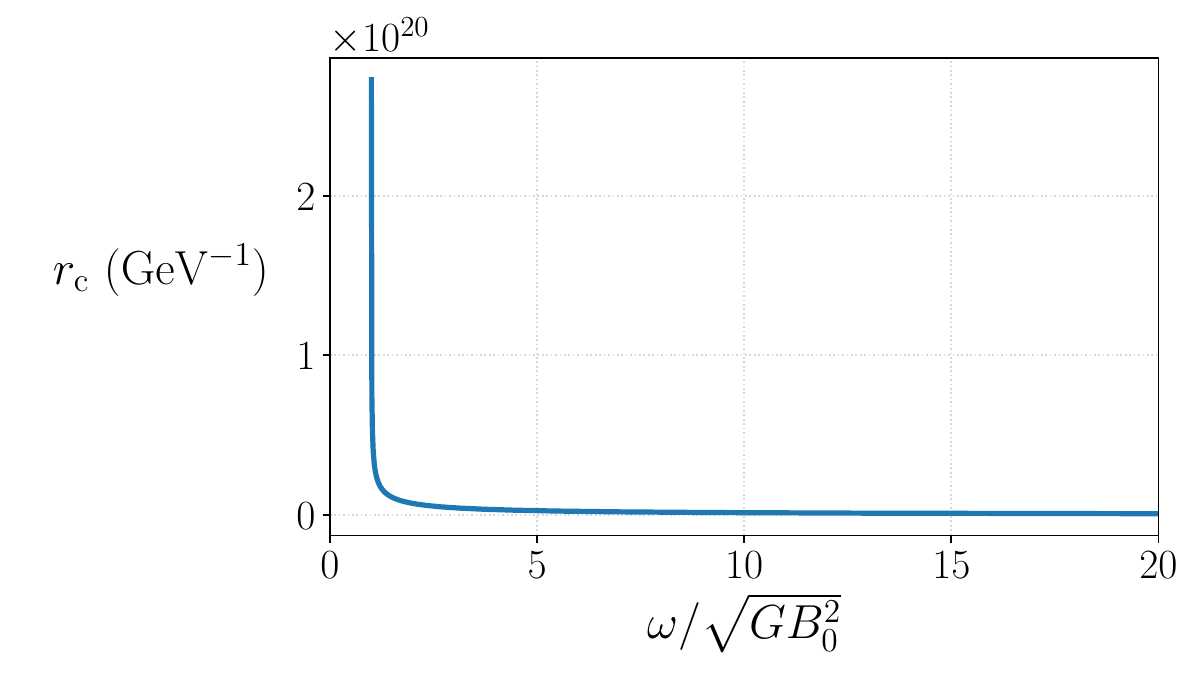}
        \caption{Behavior of the critical radius as a function of angular velocity for a magnetic field of $0.1\ \mathrm{GeV}^2$.}
        \label{fig:rcriticalvsomega-planckscalea}
    \end{subfigure}
    \hfill
    \begin{subfigure}{0.482\textwidth}
        \centering
        \includegraphics[width=\linewidth]{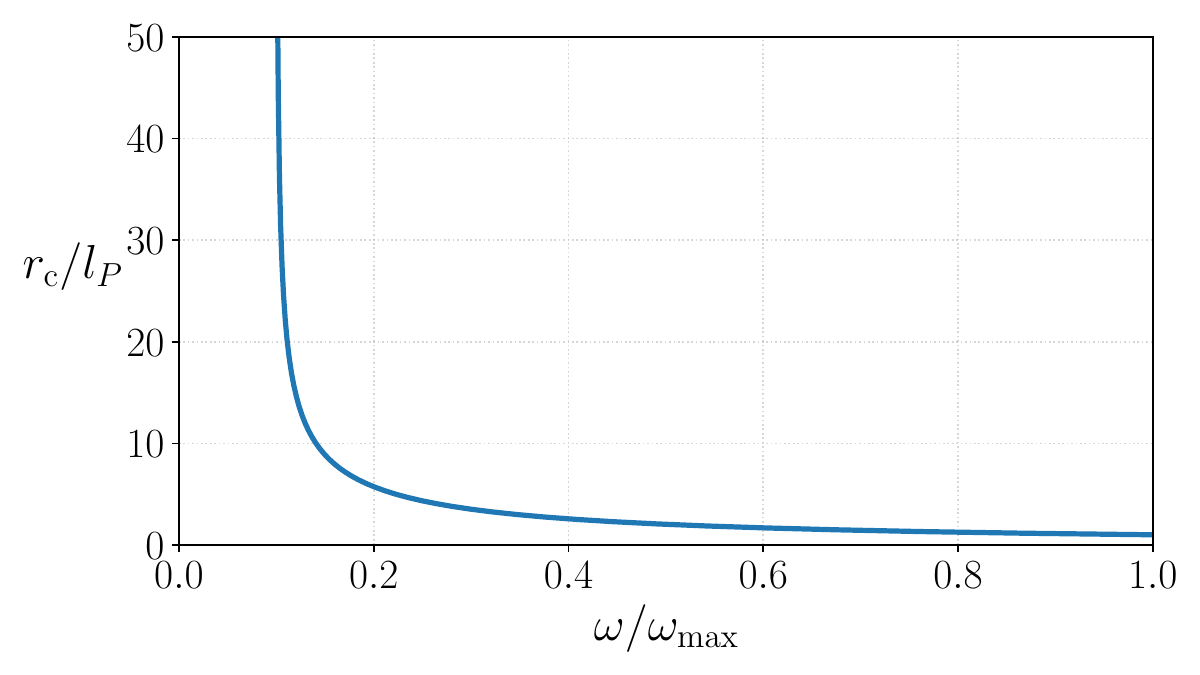}
        \caption{Behavior of the critical radius as a function of angular velocity for a magnetic field of $0.1m_P^2$.}
        \label{fig:rcriticalvsomega-planckscaleb}
    \end{subfigure}
    \caption{Comparison of critical radius behavior with respect to angular velocity for two different magnetic fields and $\omega$ scales.}
    \label{fig:combined-rcriticalvsOmega}
\end{figure}

}

\section{Klein-Gordon Equation in the Melvin Universe}\label{Klein_Gordon_Melvin}

{\color{black} In Sec. II and III we studied the main features of the Melvin spacetime and incorporated rotation in the metric. In this work, we are interested in investigating the description of} a charged scalar boson in this kind of spacetime, hence, we {\color{black} will start by considering} the Klein-Gordon equation in its covariant form
\begin{equation}
    \frac{1}{\sqrt{-g}}D_{\mu}\left(\sqrt{-g}g^{\mu\nu}D_{\nu}\Psi\right)-m^{2}\Psi=0,
\end{equation}
where \(\Psi=\Psi\left(t,r,\phi,z\right)\) is the scalar field, \(D_{\mu}=\partial_{\mu}-ieA_{\mu}\) is the derivative operator characterizing minimal coupling, \(e\) is the charge of the field, and \(A_{\mu}\) is the electromagnetic potential. {\color{black} We will first analyze the case without rotation (inertial frame), and in the next section, we consider the non-inertial effects. This way, we begin by} taking into account the Melvin metric {\color{black}from \cref{melvinfinal} and the electromagnetic potential from \cref{afinal}}. Note that the metric determinant and coefficients depend only on the radial coordinate, thus {\color{black}the resulting equation will be given by:}
\begin{equation}
    g^{tt}\frac{\partial^{2}\Psi}{\partial t^{2}}+\frac{1}{\sqrt{-g}}\frac{\partial}{\partial r}\left(\sqrt{-g}g^{rr}\frac{\partial\Psi}{\partial r}\right)+g^{\phi\phi}\left(\frac{\partial}{\partial\phi}-ieA_{\phi}\right)^{2}\Psi+g^{zz}\frac{\partial^{2}\Psi}{\partial z^{2}}-m^{2}\Psi=0.
\end{equation}
Substituting {\color{black}the expressions for} the determinant and the metric coefficients {\color{black}in the equation above}, we obtain
\begin{equation} \label{klein1}
   -\frac{1}{\Lambda^{2}}\frac{\partial^{2}\Psi}{\partial t^{2}}+\frac{1}{\Lambda^{2}r}\frac{\partial}{\partial r}\left(r\frac{\partial\Psi}{\partial r}\right)+\frac{\Lambda^{2}}{r^{2}}\left(\frac{\partial}{\partial\phi}-ieA_{\phi}\right)^{2}\Psi+\frac{1}{\Lambda^{2}}\frac{\partial^{2}\Psi}{\partial z^{2}}-m^{2}\Psi=0,
\end{equation}
and considering the cylindrical symmetry of the system, we can adopt {\color{black}an ansatz for the wave function that satisfies \cref{klein1}:}
\begin{equation}\label{Ansatz}
    \Psi\left(t,r,\phi,z\right)=e^{-i\varepsilon t}e^{i\ell\phi}e^{ip_{z}z}R\left(r\right),
\end{equation}
which leads to the following {\color{black}differential equation for the radial part of the wave function}
\begin{equation}
\frac{1}{r}\frac{d}{dr}\left(r\frac{dR\left(r\right)}{dr}\right)+\left(\varepsilon^{2}-p_{z}^{2}-\frac{\Lambda^{4}\ell^{2}}{r^{2}}+e\ell B_{0}\Lambda^{3}-\left(\frac{e^{2}B_{0}^{2}r^{2}}{4}+m^{2}\right)\Lambda^{2}\right)R\left(r\right)=0.
\end{equation}

\textcolor{black}{By employing the approximation for $\Lambda^{N}(r)$ as given in \cref{AproxLambda}, the radial equation simplifies to}
\begin{equation}
    \frac{d^{2}R\left(r\right)}{dr^{2}}+\frac{1}{r}\frac{dR\left(r\right)}{dr}+\left(\xi^{2}-\frac{\ell^{2}}{r^{2}}-\zeta^{2}r^{2}\right)R\left(r\right)=0,
\end{equation}
where we define the auxiliary quantities
\begin{equation}
    \xi^{2}=\varepsilon^{2}-p_{z}^{2}-m^{2}-GB_{0}^{2}\ell^{2}+e\ell B_{0},
\end{equation}
\begin{equation}
    \zeta^{2}=\left(\frac{e^{2}}{2}+Gm^{2}\right)\frac{1}{2}B_{0}^{2}.
\end{equation}

In order to solve the radial equation, we redefine the radial function as 
\begin{equation}
    R\left(r\right)=\frac{1}{r}u\left(r\right),
\end{equation}
which implies
\begin{equation}
    \frac{d^{2}u\left(r\right)}{dr^{2}}-\frac{1}{r}\frac{du\left(r\right)}{dr}+\left(\xi^{2}+\frac{1-\ell^{2}}{r^{2}}-\zeta^{2}r^{2}\right)u\left(r\right)=0.
\end{equation}
Now, performing the change of variable \(z=\zeta r^{2}\), the radial equation takes the form of the Whittaker equation:
\begin{equation}
    \frac{d^{2}u}{dz^{2}}+\left(-\frac{1}{4}+\frac{\kappa}{z}+\frac{\left(\frac{1}{4}-\mu^{2}\right)}{z^{2}}\right)u=0,
\end{equation}
where we define
\begin{equation}
    \kappa=\frac{\xi^{2}}{4\zeta}, \quad \mu^{2}=\frac{\ell^{2}}{4}.
\end{equation}

The general solution of the Whittaker equation is given by
\begin{equation}
u\left(z\right)=c_{1}M_{\kappa,\mu}\left(z\right)+c_{2}W_{\kappa,\mu}\left(z\right),
\end{equation}
where \(c_1\) and \(c_2\) are integration constants and \(M_{\kappa,\mu}\left(z\right)\) and \(W_{\kappa,\mu}\left(z\right)\) are the Whittaker functions of the first and second kind, respectively, explicitly given by
\begin{equation}
    M_{\kappa,\mu}\left(z\right)=e^{-\frac{1}{2}z}z^{\frac{1}{2}+\mu}M\left(\frac{1}{2}+\mu-\kappa,1+2\mu,z\right),
\end{equation}
\begin{equation}
    W_{\kappa,\mu}\left(z\right)=e^{-\frac{1}{2}z}z^{\frac{1}{2}+\mu}U\left(\frac{1}{2}+\mu-\kappa,1+2\mu,z\right),
\end{equation}
with \(M\left(a,b,z\right)\) and \(U\left(a,b,z\right)\) being Kummer's confluent hypergeometric functions of the first and second kind, respectively. Since we are interested in a solution that is well-behaved at the origin, we set \(c_2=0\), therefore the radial solution becomes
\begin{equation}
  R\left(r\right)\sim\frac{1}{r}e^{-\frac{1}{2}\zeta r^{2}}r^{1+\left|\ell\right|}M\left(\frac{1}{2}+\frac{\left|\ell\right|}{2}-\frac{\xi^{2}}{4\zeta},1+\left|\ell\right|,\zeta r^{2}\right).
\end{equation}

We require a solution that is well-behaved both at the origin and at infinity; thus, we impose the condition for a polynomial solution of \(M\left(a,b,z\right)\), which is given by
\begin{equation}
    a=-n,\quad n=0,1,2,\ldots.
\end{equation}

By substituting the auxiliary quantities and performing the necessary manipulations, we obtain the energy spectrum associated with the scalar field in the Melvin universe:
\begin{equation}\label{eq:espectro}
    \varepsilon_{\pm} = \pm\sqrt{m^{2}+p_{z}^{2}+GB_{0}^{2}\ell^{2}-e\ell B_{0}+B_{0}\sqrt{e^{2}+2Gm^{2}}\left(2n+1+\left|\ell\right|\right)}.
\end{equation}

In the particular case where the magnetic field intensity tends to zero, we recover the flat spacetime energy spectrum,
\begin{equation}
    \varepsilon_{\pm}=\pm\sqrt{m^{2}+p_{z}^{2}}.
\end{equation}

An interesting regime is obtained by considering the case \(Gm^{2}\approx0\), so that an expansion in terms involving the square root of the product of the gravitational constant and the mass yields
\begin{equation}
\varepsilon_{\pm}=\pm\sqrt{m^{2}+p_{z}^{2}+GB_{0}^{2}\ell^{2}+e^{-1}m^{2}GB_{0}\left(2n+1+\left|\ell\right|\right)+eB_{0}\left(2n+1+\left|\ell\right|-\ell\right)}.
\end{equation}

It should be noted that the spectra contain terms proportional to \(eB_0n\) and \(e^{-1}m^{2}GB_{0}n\); the former are well-known from the study of Landau levels, while the latter exhibit dependencies on the gravitational constant, which may be interpreted as gravitational corrections to the Landau levels.

{\color{black}
The behavior of the radial wave function is depicted in \autoref{fig:radialwavefunction-variedn-1gev2}, for a particle with mass $m_\pi\approx135\ \mathrm{MeV}$ and a magnetic field of $1\ \mathrm{GeV}^2$, which shows the characteristic radial distance of the system to be within 1 fm, and in \autoref{fig:radialwavefunction-variedn}, for a much higher magnetic field ($0.001m_P^2$), up to the Planck scale, which we can identify as the system being mostly localized up to 200 Planck lengths. In both cases, one can observe the oscillatory character with the wave function approaching zero as $r$ increases for $n=\{0,1,2,3\}$. Conversely, the respective radial probability density in \autoref{fig:combined-probdens} illustrates that the number of peaks for a given state with quantum number $n$ is $n+1$.

\begin{figure}[h!] \label{fig:radialwavefunction3}
    \centering
    \begin{subfigure}{0.464\textwidth}
        \centering
        \includegraphics[width=\linewidth]{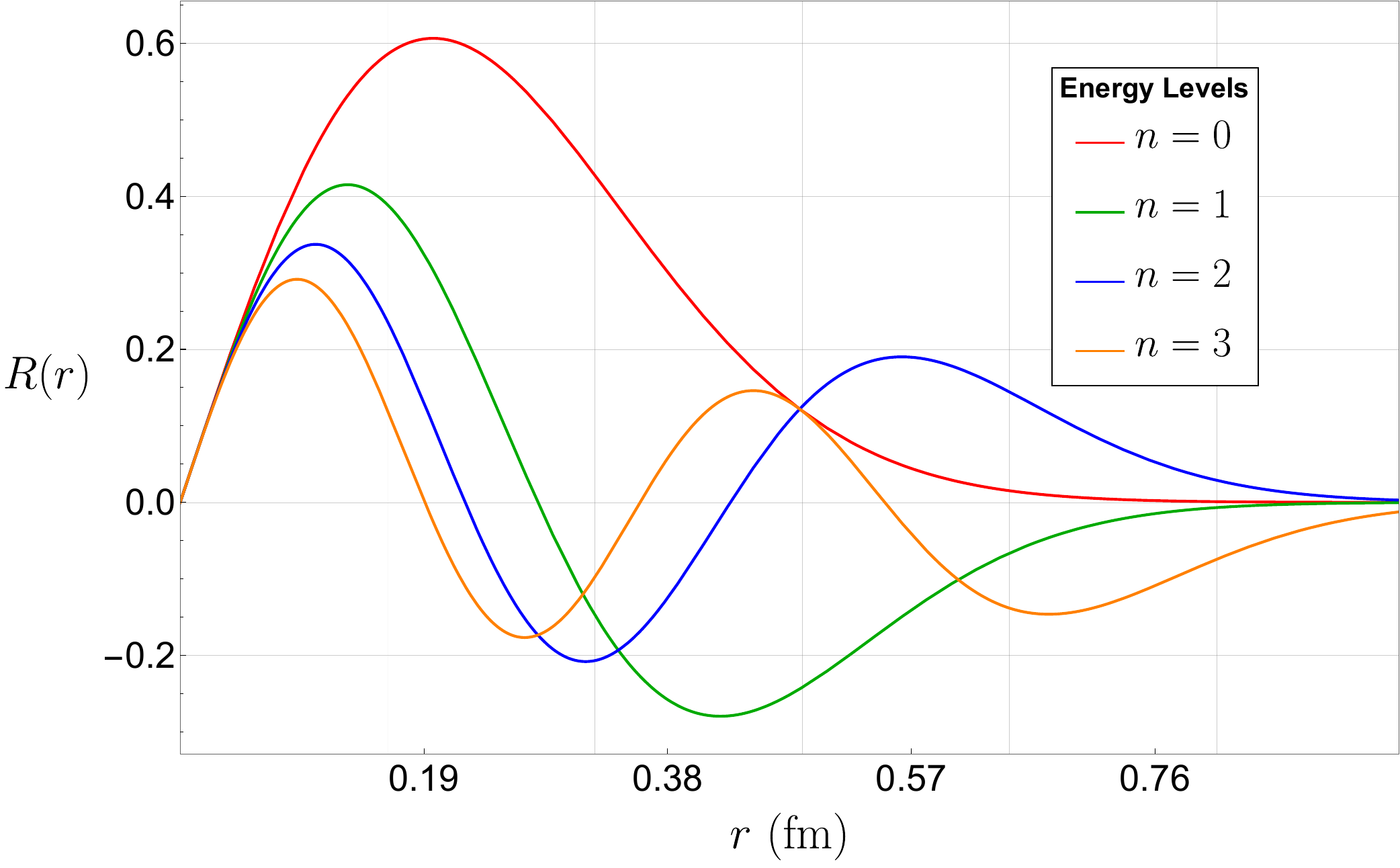}
        \caption{Plot of $R(r)$ against $r$, with parameters $m=m_\pi$, $e=q_e$, $B_0=1\ \mathrm{GeV}^2$, $p_z=0$ and quantum numbers $\ell=1$ and $n=\{0,1,2,3\}$.}
        \label{fig:radialwavefunction-variedn-1gev2}
    \end{subfigure}
    \hfill
    \begin{subfigure}{0.49\textwidth}
        \centering
        \includegraphics[width=\linewidth]{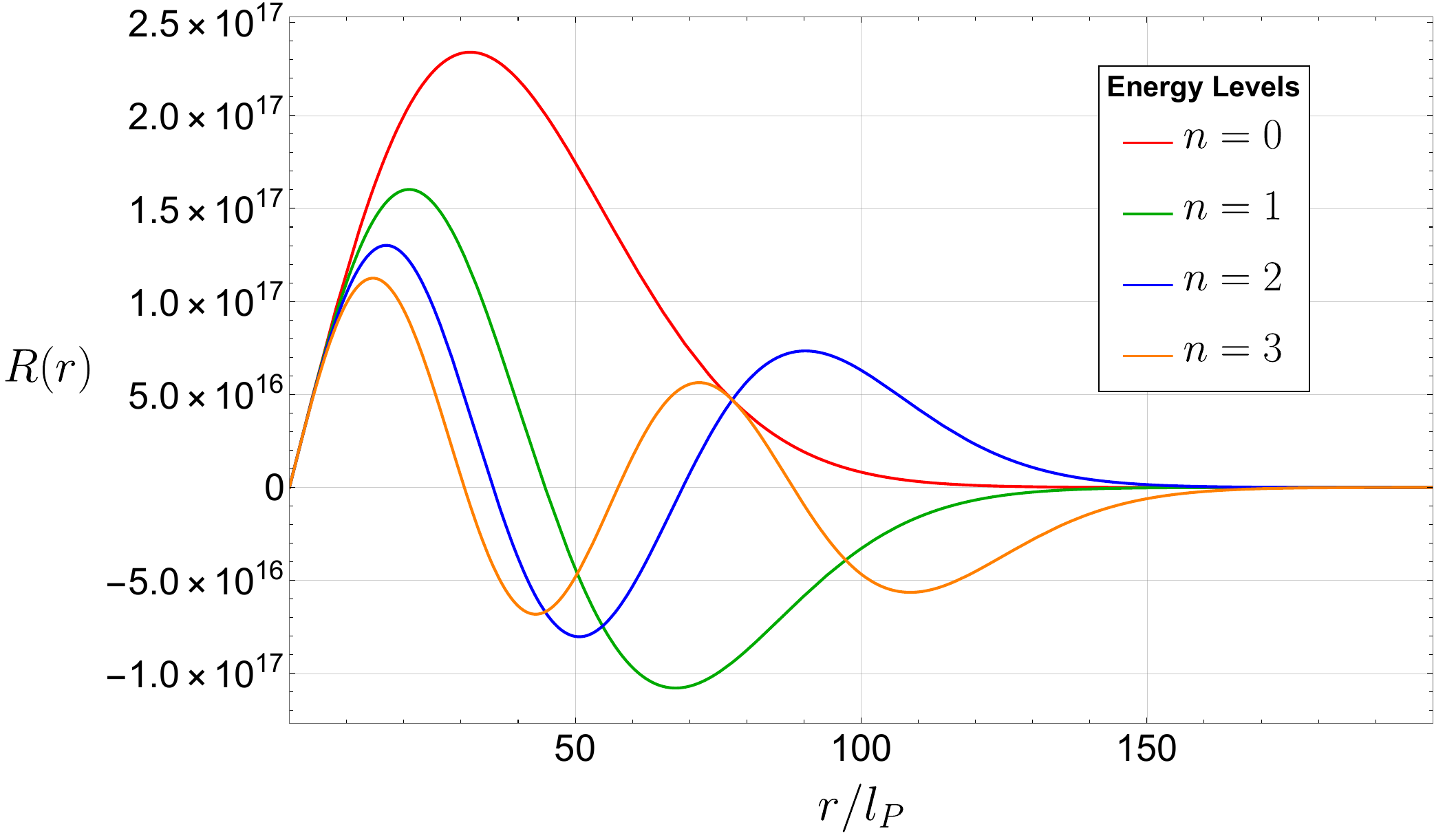}
        \caption{Plot of $R(r)$ against $r$, with parameters $m=m_\pi$, $e=q_e$, $B_0=0.001m_P^2$, $p_z=0$ and quantum numbers $\ell=1$ and $n=\{0,1,2,3\}$.}
        \label{fig:radialwavefunction-variedn}
    \end{subfigure}
    \caption{Comparison of radial wave functions for different magnetic field strengths.}
    \label{fig:combined}
\end{figure}

\begin{figure}[H]
    \centering
    \begin{subfigure}{0.464\textwidth}
        \centering
        \includegraphics[width=\linewidth]{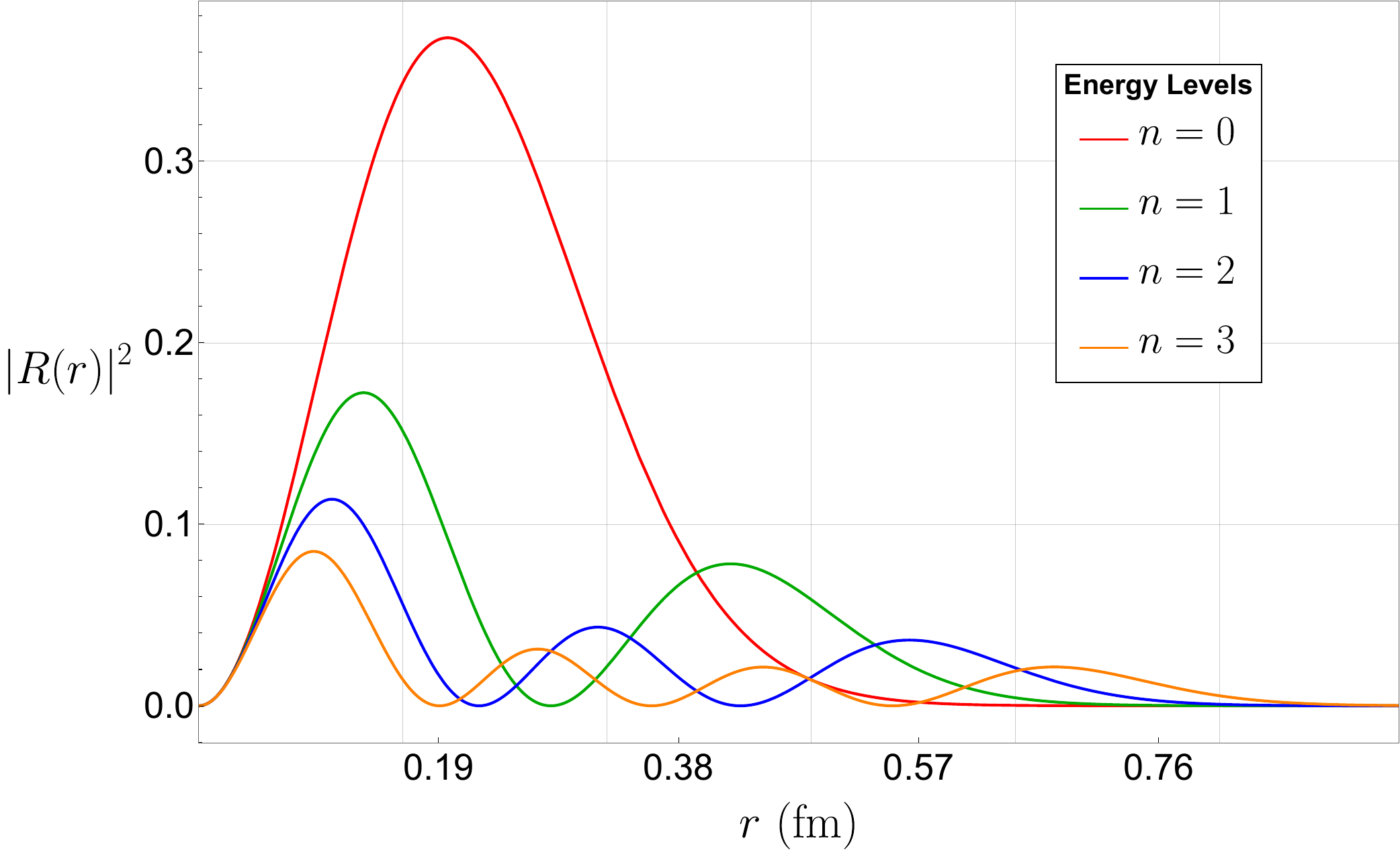}
        \caption{Plot of $|R(r)|^2$ against $r$, with parameters $m=m_\pi$, $e=q_e$, $B_0=1\ \mathrm{GeV}^2$, $p_z=0$, and quantum numbers $\ell=1$ and $n=\{0,1,2,3\}$.}
        \label{fig:probdenswavefunction-variedn-1gev2}
    \end{subfigure}
    \hfill
    \begin{subfigure}{0.49\textwidth}
        \centering
        \includegraphics[width=\linewidth]{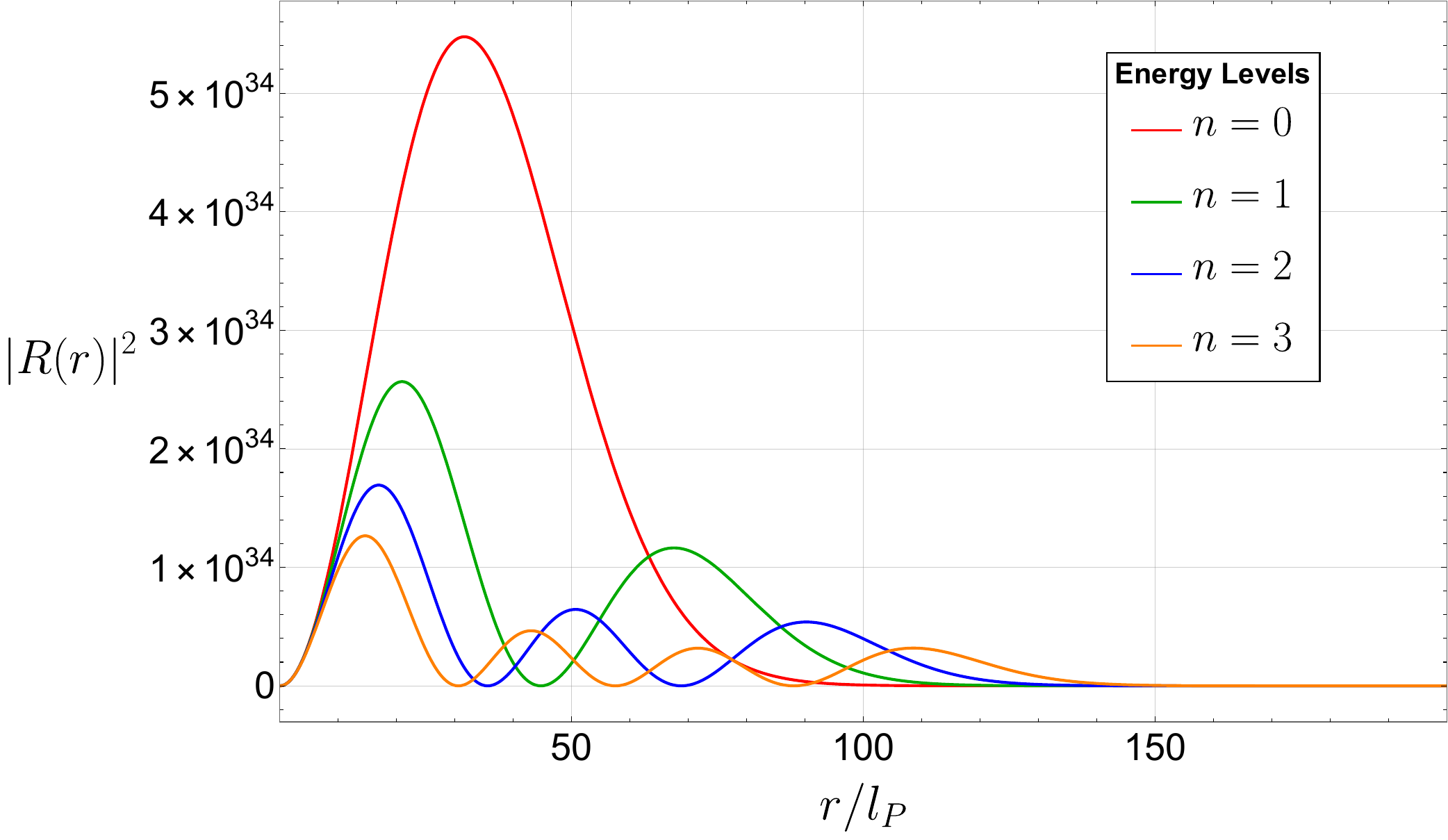}
        \caption{Plot of $|R(r)|^2$ against $r$, with parameters $m=m_\pi$, $e=q_e$, $B_0=0.1m_P^2$, $p_z=0$, and quantum numbers $\ell=1$ and $n=\{0,1,2,3\}$.}
        \label{fig:probdenswavefunction-variedn}
    \end{subfigure}
    \caption{Probability density functions for different magnetic field strengths.}
    \label{fig:combined-probdens}
\end{figure}

The solution's magnetic field dependence, as evidenced in previous figures, delocalizes the wave function towards the origin. Specifically, \autoref{fig:radialwavefunction-variedB0-1gev2} shows that for the lowest energy state with non-null angular momentum ($n=0$, $\ell=1$) coupling with $B_0^2$, the magnetic field attracts the system towards the origin. Consequently, we expect the average radial distance to decrease with increasing $B_0$.

\begin{figure}[H]
    \centering
    \includegraphics[width=0.7\linewidth]{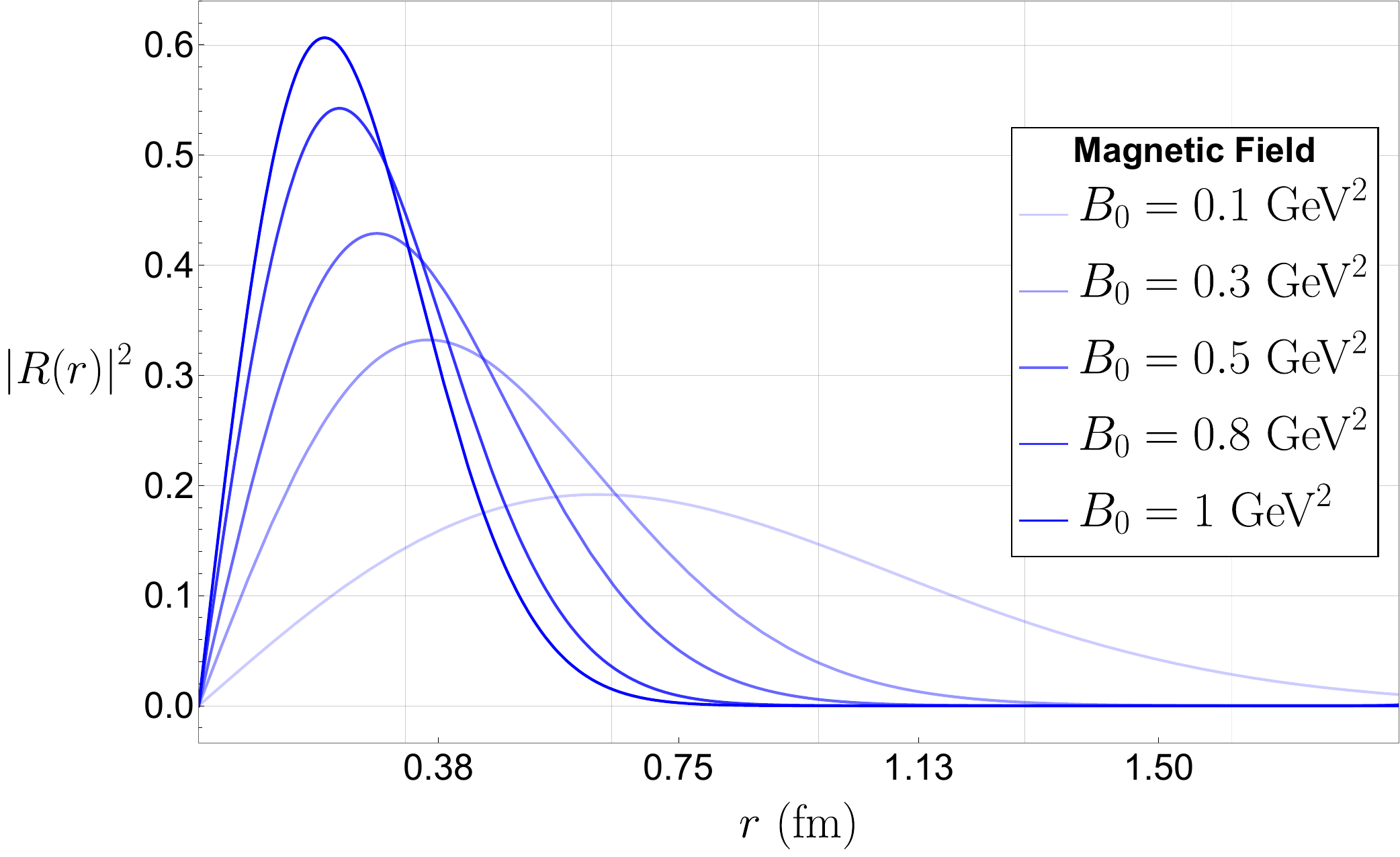}
    \caption{Plot of $R(r)$ against $r$ for several values of magnetic fields, with parameters $m=m_\pi$, $e=q_e$, $p_z=0$ and quantum numbers $\ell=1$, and $n=0$.}
    \label{fig:radialwavefunction-variedB0-1gev2}
\end{figure}
}

\begin{figure}[!ht]
    \centering
    \begin{subfigure}{0.49\textwidth}
        \centering
        \includegraphics[width=\linewidth]{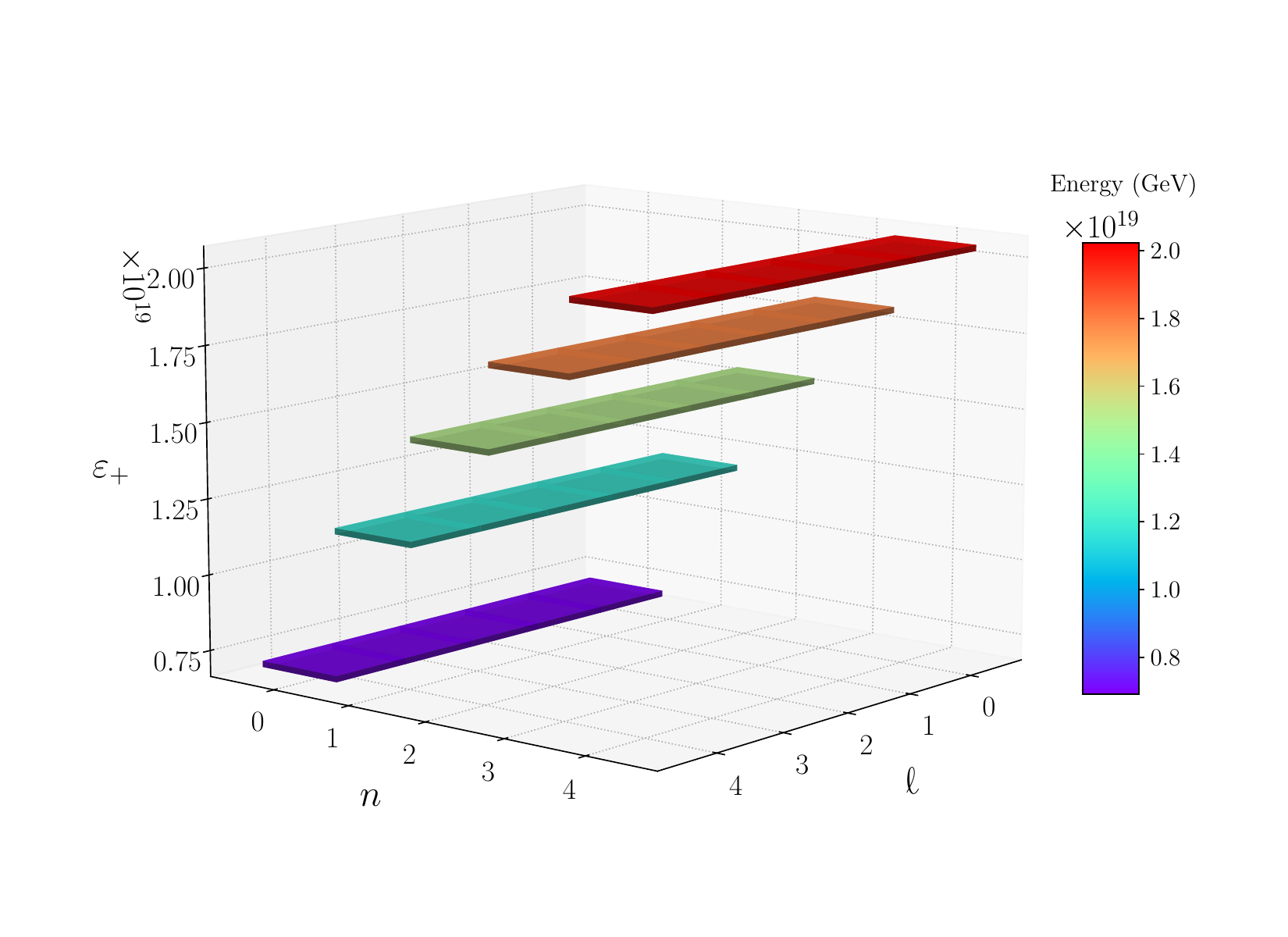}
        \caption{Positive energy levels against quantum numbers $n$ and $\ell$, with parameters $m = m_\pi$, $e = q_e$, $B_0 = 1\,\mathrm{GeV}^2$, $p_z = 0$.}
        \label{fig:discrete3dnl-1gev2}
    \end{subfigure}
    \hfill
    \begin{subfigure}{0.49\textwidth}
        \centering
        \includegraphics[width=\linewidth]{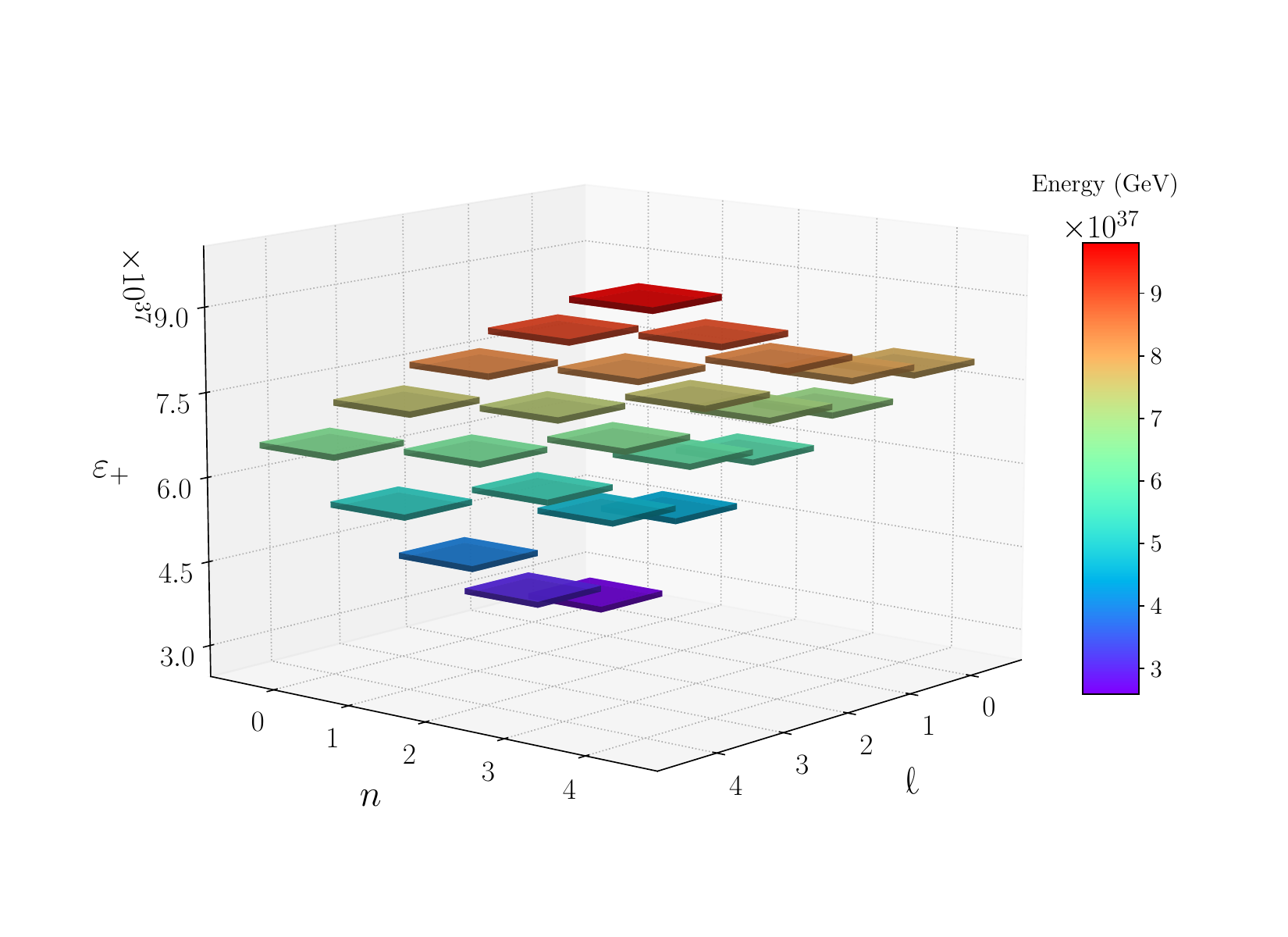}
        \caption{Positive energy levels against quantum numbers $n$ and $\ell$, with parameters $m = m_\pi$, $e = q_e$, $B_0 = 0.1\,m_P^2$, $p_z = 0$.}
        \label{fig:discrete3dnl}
    \end{subfigure}
    \caption{Comparison of positive energy levels for different magnetic field strengths.}
    \label{fig:combined-discrete3dnl}
\end{figure}

{\color{black} The energy spectrum's dependence on the principal quantum number $n$ and angular momentum quantum number $\ell$ reveals gravitational contributions near the Planck scale (see \cref{eq:espectro}). Comparing \autoref{fig:discrete3dnl-1gev2} and \autoref{fig:discrete3dnl} illustrates this: at the heavy-ion collision scale, flat spacetime suppresses gravitational corrections, but with a magnetic field of $0.1m_P^2$, these corrections become significant, noticeably shifting the dependence on $\ell$ for each $n$. Furthermore, we can see how larger values of $\ell$ significantly increase the energy level. }

{\color{black} 

Additionally, we can refer to \cref{eq:espectro} to argue that the magnetic field has a greater impact on quantum states with non-null angular momentum quantum numbers ($\ell \neq 0$), and as expected, the scale at which the energy shift due to such a field is non-negligible is for $B_0$ at orders of $m_P^2 \left(\sim 10^{38}\ \mathrm{GeV}^2\right)$ in natural units, representing around $10^{53}\ \mathrm{T}$ or $10^{57}\ \mathrm{G}$, in SI. But, independent of these very large magnetic fields, the scalar particle can have its energy altered by fields below $10^{19}\ \mathrm{G}$, as we can see from \autoref{fig:energyagainstb0}, where it is possible to notice, for example, that around $B_0\sim0.6\ \mathrm{GeV}^2$ the energy can increase near five times compared with the scalar particle's rest mass energy.

\begin{figure}[H]
    \centering
    \includegraphics[width=0.7\linewidth]{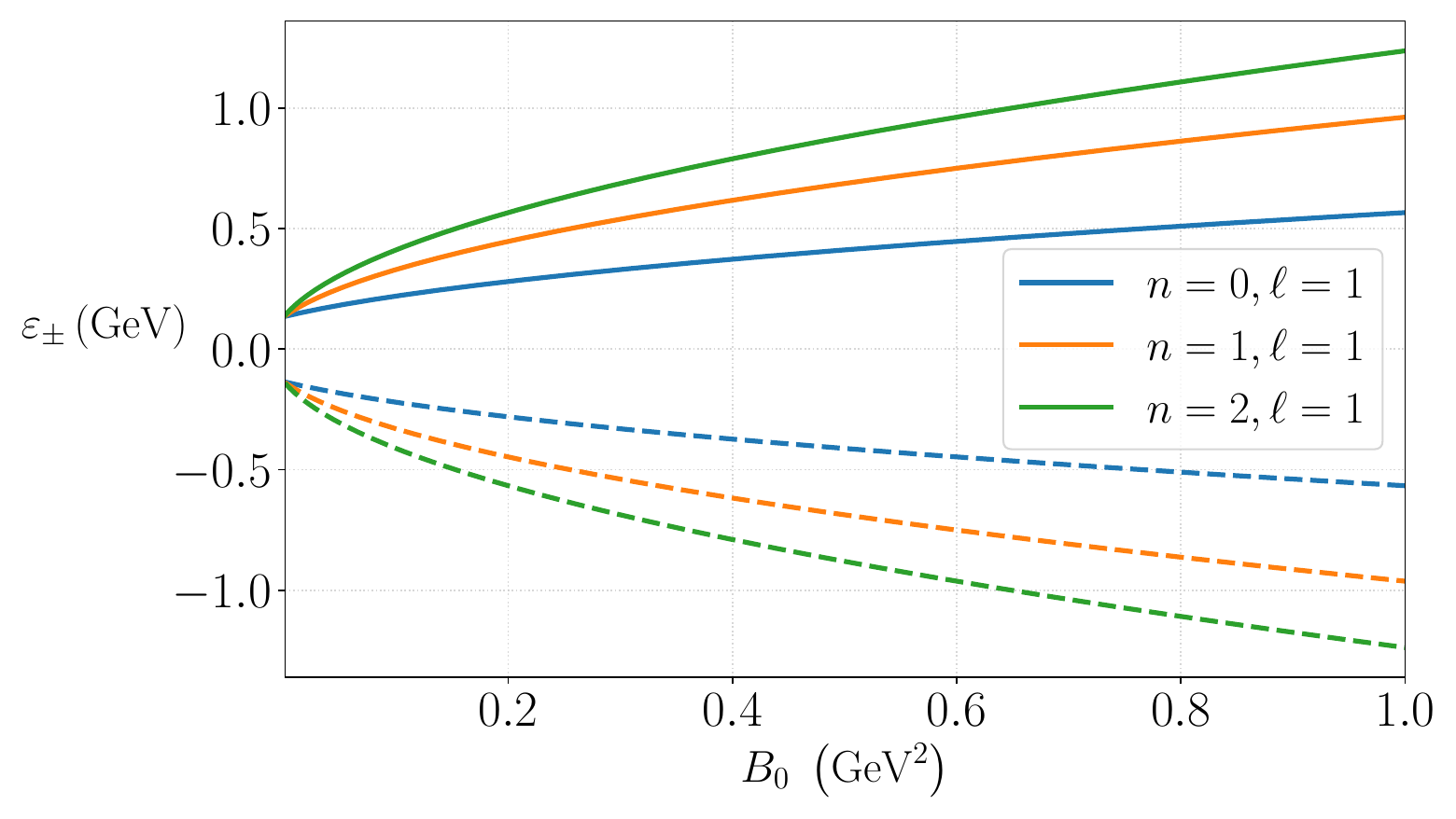}
    \caption{Energy levels against $B_0$, with parameters $m=m_\pi$, $e=q_e$, $p_z=0$ and quantum numbers $\ell=1$, $n=\{0,1,2\}$. Positive energies $\varepsilon_+$ are shown in solid lines and negative energies $\varepsilon_-$ in dashed ones.}
    \label{fig:energyagainstb0}
\end{figure}

We can numerically compare the value of energy with magnetic fields of the order as those found in magnetars ($B\sim10^{16}\ \mathrm{G}$) or in heavy ion collisions ($B\sim10^{19}\ \mathrm{G}$), and for that we refer to \autoref{tab:tabeladeltaenergyNatural} which we can notice a notable energy shift for quantum numbers $n$ as large as $10^{36}$ similarly to \cite{santos} for the highest magnetic field intensities known.

\begin{table}[H]
\centering
\caption{Table with values of $\Delta \varepsilon = \varepsilon_{+}-m$ for different values of magnetic fields and quantum numbers $n$, with parameters $m=m_\pi$, $e=q_e$, $p_z=0$, and $\ell=1$.}
\begin{tabular}{l c c c}
\toprule[1.5pt]
 & \( n=10^{25} \) (GeV) & \( n=10^{30} \) (GeV) & \( n=10^{36} \) (GeV) \\
\midrule[1.5pt]
\( B_{\text{Earth}} \sim 0.56 \, \text{G} \) & 257.24958 & $8.139 \times 10^{4}$ & $8.139 \times 10^{7}$ \\
\( B_{\text{Sun}} \sim 1.5 \, \text{G} \) & 421.10936 & $1.332 \times 10^{5}$ & $1.332 \times 10^{8}$ \\
\( B_{\text{Pulsar}} \sim 10^{13} \, \text{G} \) & $1.088 \times 10^{9}$ & $3.439 \times 10^{11}$ & $3.439 \times 10^{14}$ \\
\( B_{\text{Magnetar}} \sim 10^{15} \, \text{G} \) & $1.088 \times 10^{10}$ & $3.439 \times 10^{12}$ & $3.439 \times 10^{15}$ \\
\( B_{\text{Ions}} \sim 10^{19} \, \text{G} \) & $1.088 \times 10^{12}$ & $3.439 \times 10^{14}$ & $3.439 \times 10^{17}$ \\
\bottomrule[1.5pt]
\end{tabular}
\label{tab:tabeladeltaenergyNatural}
\end{table}

We can also perceive this energy shift by visualizing \autoref{fig:energyvsn}, where we make a similar analysis to \cite{santos}.

\begin{figure}[H]
    \centering
    \includegraphics[width=0.7\linewidth]{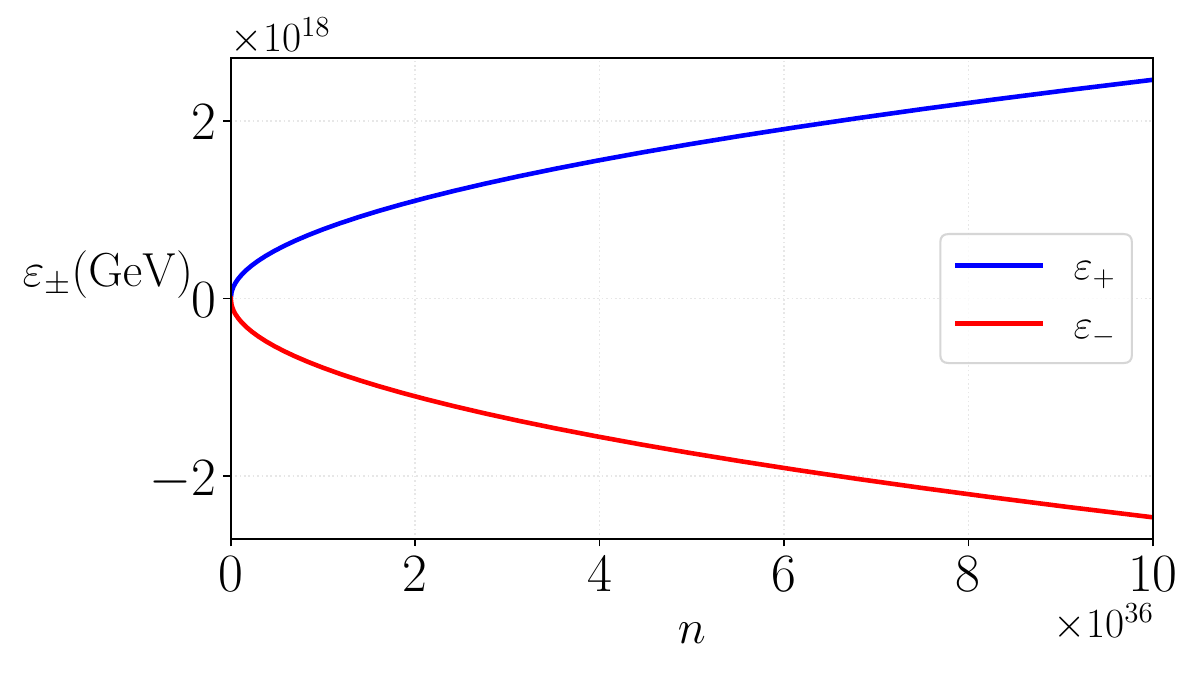}
    \caption{Energy levels against quantum number $n$, with parameters $m=m_\pi$, $e=1$, $B_0=1\ \mathrm{GeV^2} \sim10^{19}\ \mathrm{G}$, $p_z=0$, and $\ell=1$.}
    \label{fig:energyvsn}
\end{figure}

}

\section{Klein-Gordon Equation in the Melvin Universe with Non-inertial Effects}\label{Klein_Gordon_Non_Inertial_Melvin}
\textcolor{black}{By introducing non-inertial effects through a rotating reference frame, the structure of the Melvin spacetime is modified, and the corresponding Klein-Gordon equation for a charged scalar field acquires the form:}
\begin{align}
&\frac{1}{\sqrt{-g}} \frac{\partial}{\partial r} \left( \sqrt{-g}\, g^{rr} \frac{\partial \Psi}{\partial r} \right)
+ g^{tt} \left( \frac{\partial}{\partial t} - i e A_{t} \right)^{2} \Psi \notag \\
&\quad + g^{\chi\chi} \left( \frac{\partial}{\partial \chi} - i e A_{\chi} \right)^{2} \Psi
+ 2 g^{t\chi} \left( \frac{\partial}{\partial t} - i e A_{t} \right) \left( \frac{\partial}{\partial \chi} - i e A_{\chi} \right) \Psi \notag \\
&\quad + g^{zz} \frac{\partial^{2} \Psi}{\partial z^{2}} - m^{2} \Psi = 0.
\end{align}

\textcolor{black}{After substituting the metric determinant and its components, and performing some algebraic manipulations, we obtain:}
\begin{equation}
\begin{split}
&\frac{1}{\Lambda^{2} r} \frac{\partial}{\partial r} \left( r \frac{\partial \Psi}{\partial r} \right)
- \frac{1}{\Lambda^{2}} \left( \frac{\partial}{\partial t} - i e A_{t}(r) \right)^{2} \Psi \\
&\quad + \left( \frac{\Lambda^{2}}{r^{2}} - \frac{\omega^{2}}{\Lambda^{2}} \right) \left( \frac{\partial}{\partial \chi} - i e A_{\chi} \right)^{2} \Psi
+ \frac{2 \omega}{\Lambda^{2}} \left( \frac{\partial}{\partial t} - i e A_{t} \right) \left( \frac{\partial}{\partial \chi} - i e A_{\chi} \right) \Psi \\
&\quad + \frac{1}{\Lambda^{2}} \frac{\partial^{2} \Psi}{\partial z^{2}} - m^{2} \Psi = 0.
\end{split}
\end{equation}

Analogously to the previous case, by taking the ansatz from \cref{Ansatz} for the wave function, we obtain the radial equation for the wave function
\begin{equation}
    \begin{split}
       &\frac{1}{r}\frac{d}{d r}\left(r\frac{d R(r)}{d r}\right)+\Biggl(\left(\varepsilon+\omega\ell\right)^{2}-p_{z}^{2}-\left(m^{2}+\frac{e^{2}B_{0}^{2}r^{2}}{4}\right)\Lambda^{2}+\ell eB_{0}\Lambda^{3}-\frac{\ell^{2}}{r^{2}}\Lambda^{4}\Biggr)R(r)\\
       &\quad+\underbrace{\left(\frac{\omega\varepsilon eB_{0}r^{2}}{\Lambda}-\frac{\omega\varepsilon eB_{0}r^{2}}{\Lambda}+\frac{\omega^{2}\ell eB_{0}r^{2}}{\Lambda}-\frac{\omega^{2}\ell eB_{0}r^{2}}{\Lambda}\right)}_{=0}R(r)\\
       &\quad+\underbrace{\left(\frac{\omega^{2}e^{2}B_{0}^{2}r^{4}}{2\Lambda^{2}}-\frac{\omega^{2}e^{2}B_{0}^{2}r^{4}}{2\Lambda^{2}}\right)}_{=0}R(r)=0.
    \end{split}
\end{equation}

Canceling the appropriate terms, we obtain the following radial equation for the system with non-inertial effects  
\begin{equation}
  \frac{1}{r}\frac{d}{dr}\left(r\frac{dR(r)}{dr}\right)+\left(\left(\varepsilon+\omega\ell\right)^{2}-p_{z}^{2}-\frac{\Lambda^{4}\ell^{2}}{r^{2}}+e\ell B_{0}\Lambda^{3}-\left(\frac{e^{2}B_{0}^{2}r^{2}}{4}+m^{2}\right)\Lambda^{2}\right)R(r)=0.
\end{equation}

We note that this equation is very similar to the one in the case without non-inertial effects; the only difference is that now the energy term is augmented by a contribution from the product of the angular velocity with the quantum number associated with the angular momentum, \(\omega\ell\). As done previously, considering the approximation {\color{black} for $\Lambda^{N}(r)$ given in \cref{AproxLambda}}, we obtain the radial solution for {\color{black}the wave function with non-inertial effects to be as follows}
\begin{equation}
    R(r)\sim\frac{1}{r}e^{-\frac{1}{2}\zeta r^{2}}r^{1+|\ell|}M\left(\frac{1}{2}+\frac{|\ell|}{2}-\frac{\tau^{2}}{4\zeta},1+|\ell|,\zeta r^{2}\right),
\end{equation}
where \(\tau\) {\color{black}is the parameter that carries the dependency on $\omega$, and} is given by 
\begin{equation}
    \tau^{2}=\left(\varepsilon+\omega\ell\right)^{2}-p_{z}^{2}-m^{2}-GB_{0}^{2}\ell^{2}+e\ell B_{0}. 
\end{equation}

\textcolor{black}{In the limit where $\omega \rightarrow 0$, the parameter $\tau$ reduces to $\xi$, recovering the radial equation without non-inertial effects. As far as these results depend on $\omega$, it is interesting to study some cases for different values of this parameter.}

\subsection{Case: $\omega\rightarrow\omega_{\text{min}}$}
\textcolor{black}{In the analysis of the radial function under non-inertial effects, it is necessary to consider the range defined for the critical radius (\cref{Criticalradio}). When the rotation parameter approaches its minimum value, $\omega_{\text{min}}$ (\cref{OmegaInt}), the critical radius tends to infinity. In this limit, the polynomial condition on Kummer’s confluent hypergeometric function must be imposed to ensure that the solution remains physically acceptable. This condition leads to the quantization of the system, and the corresponding energy spectrum is given by}
\begin{equation}
    \varepsilon_{\pm}=-\omega\ell\pm\sqrt{m^{2}+p_{z}^{2}+GB_{0}^{2}\ell^{2}-e\ell B_{0}+B_{0}\left(e^{2}+2Gm^{2}\right)^{\frac{1}{2}}\left(2n+1+|\ell|\right)}.
\end{equation}

We note that in the limit of zero magnetic field intensity, there is now an angular velocity contribution to the flat spacetime energy,
\begin{equation}
    \varepsilon_{\pm}=-\omega\ell\pm\sqrt{m^{2}+p_{z}^{2}}.
\end{equation}

As done previously {\color{black}for the solution in the inertial frame, we consider the regime where \(Gm^{2}\) is small, which in the present case leads to}
\begin{equation} 
 \varepsilon_{\pm}=-\omega\ell\pm\sqrt{m^{2}+p_{z}^{2}+GB_{0}^{2}\ell^{2}+e^{-1}m^{2}GB_{0}\left(2n+1+\left|\ell\right|\right)+eB_{0}\left(2n+1+\left|\ell\right|-\ell\right)},
  \label{eq:espectroNonInertial}
\end{equation}
so that both energy spectra recover the previous cases for \(\omega=0\). Thus, we can notice that the energy levels are shifted by non-inertial effects in spacetime. {\color{black}We observe that for positive $\omega$, the energy levels (\cref{eq:espectroNonInertial}) decrease, shifting the positive eigenvalues downward. Each \{$n,l$\} configuration has a specific critical frequency, beyond which the positive energies become negative. \autoref{fig:combined-discrete3dnl-noninertial} illustrates the dependence on quantum numbers $n$ and $l$ for heavy-ion collision and Planck scales, using low $\omega$ values. This figure reveals the sensitivity of energy levels with non-zero angular momentum $l$ to angular velocity, disrupting the pattern observed in \autoref{fig:combined-discrete3dnl} and resulting in negative eigenvalues.

\begin{figure}[H]
    \centering
    \begin{subfigure}{0.49\textwidth}
        \centering
        \includegraphics[width=\linewidth]{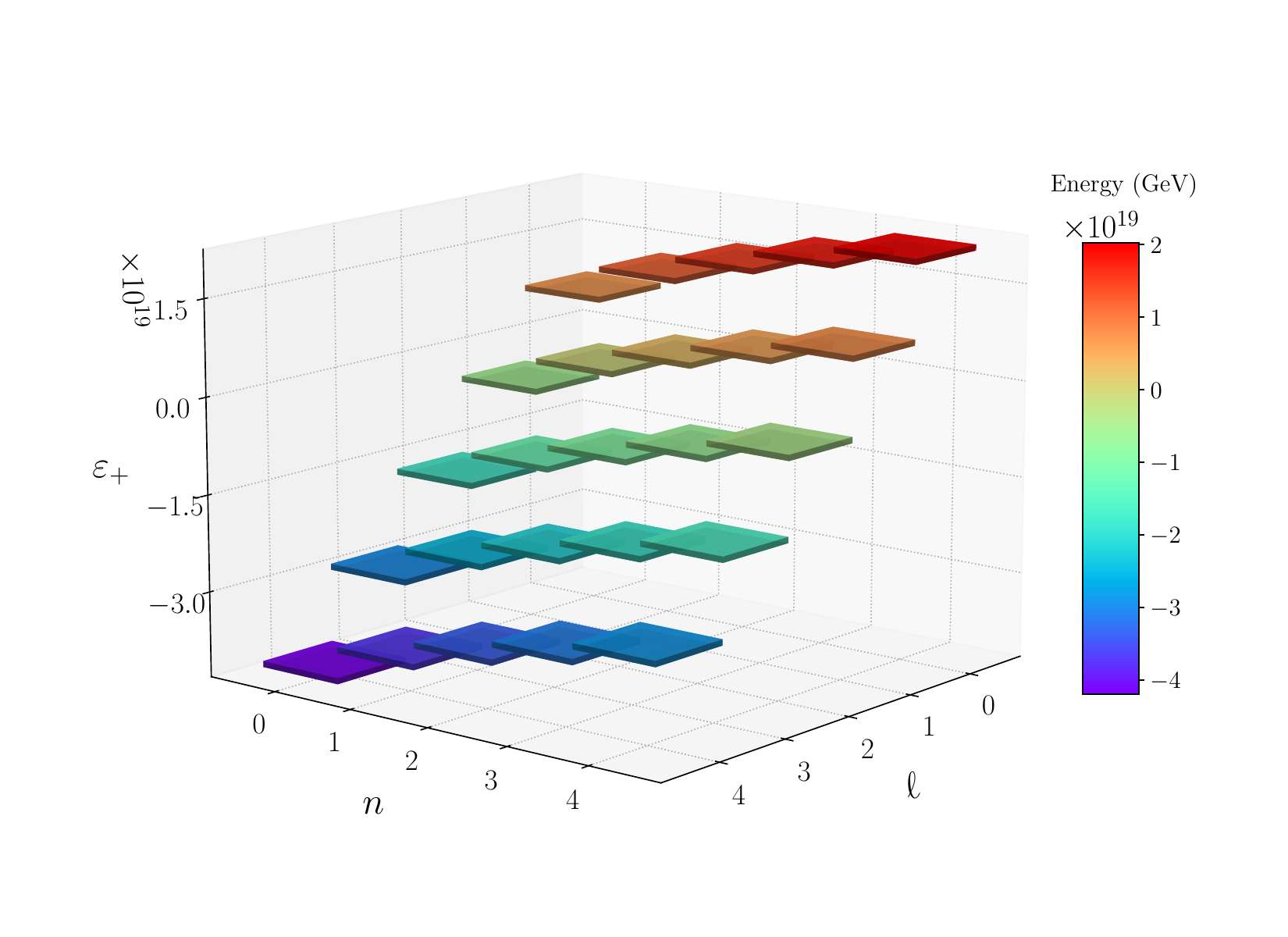}
        \caption{Positive energy levels with non-inertial effects against quantum numbers $n$ and $\ell$, for a magnetic field $B_0 = 1\,\mathrm{GeV}^2$ and $\omega=1\,\mathrm{GeV}\ (\sim10^{24}\ \mathrm{Hz})$, with parameters $m = m_\pi$, $e = q_e$, and $p_z = 0$.}
        \label{fig:discrete3dnl-1gev2-noninertial}
    \end{subfigure}
    \hfill
    \begin{subfigure}{0.49\textwidth}
        \centering
        \includegraphics[width=\linewidth]{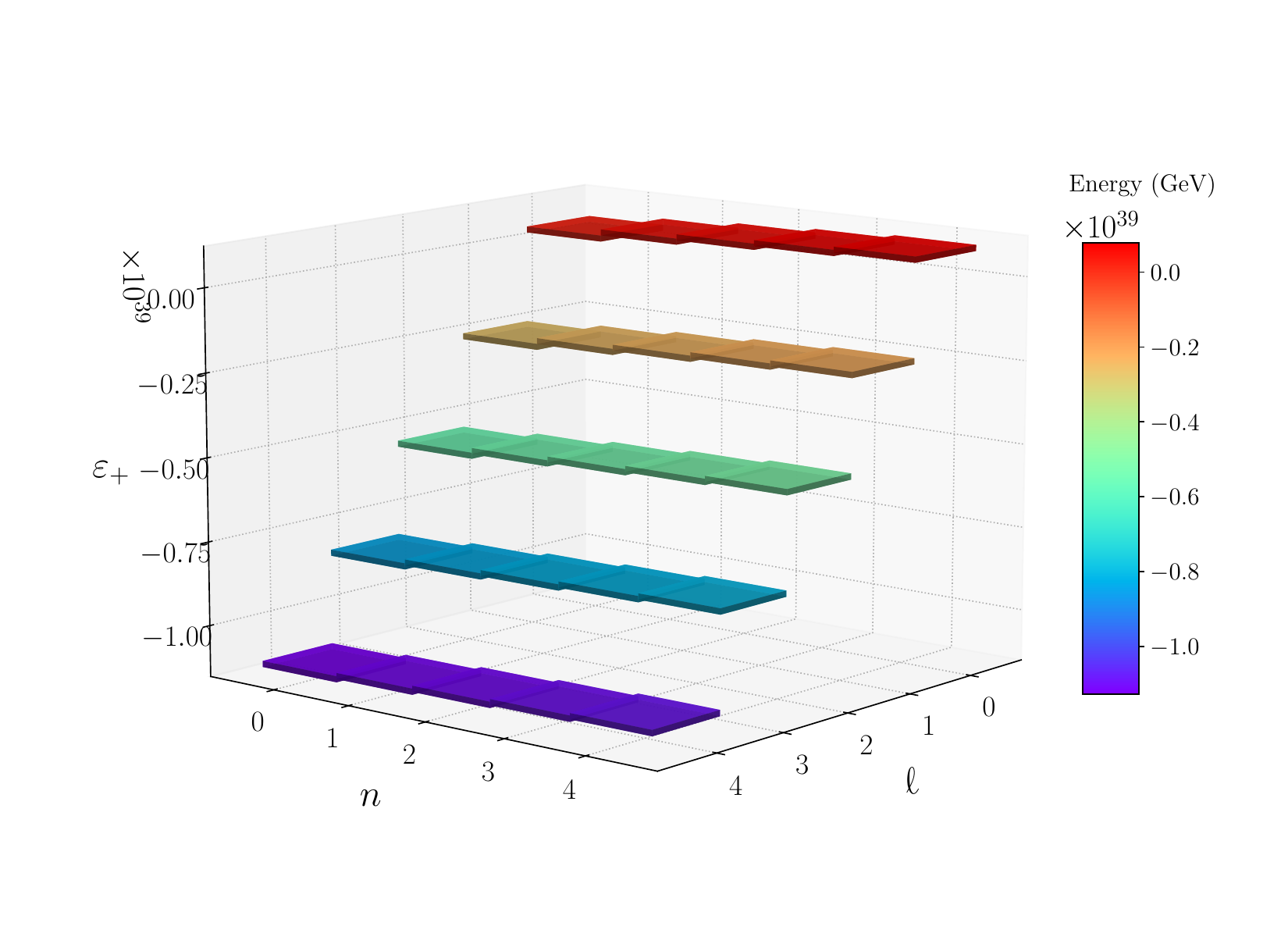}
        \caption{Positive energy levels with non-inertial effects against quantum numbers $n$ and $\ell$, for a magnetic field $B_0 = 0.1m_P^2$ and $\omega=20\sqrt{GB_0^2}\ (\sim10^{43}\ \mathrm{Hz})$, with parameters $m = m_\pi$, $e = q_e$, and $p_z = 0$.}
        \label{fig:discrete3dnl-noninertial}
    \end{subfigure}
    \caption{Comparison of positive energy levels for different magnetic field strengths with non-inertial effects in spacetime.}
    \label{fig:combined-discrete3dnl-noninertial}
\end{figure}
}

\subsection{Case: $\omega_{\text{max}}\geq\omega\gg\omega_{\text{min}}$}

\textcolor{black}{When the rotation parameter lies within the interval $\omega_{\text{max}}\geq\omega\gg\omega_{\text{min}}$, the critical radius $r_{c}$ takes finite values. In this regime, we impose the boundary condition $R(r_{c}) = 0$, commonly referred to as the hard-wall condition. For small arguments, i.e., $z_{0} \ll 1$, the confluent hypergeometric function of the first kind, $M(a,b,z_{0})$, can be approximated as \cite{abramowitz1965handbook}
\begin{equation}
M(a,b,z_{0}) \approx \frac{\Gamma(b)}{\sqrt{\pi}} e^{\frac{z_{0}}{2}} z_{0}^{\frac{1-b}{2}} \left( \frac{b}{2} - a \right)^{\frac{1-b}{2}} \cos \left( \sqrt{2z_{0}(b - 2a)} - \frac{b\pi}{2} + \frac{\pi}{4} \right),
\end{equation}
where $\Gamma(b)$ denotes the Gamma function. Assuming $\zeta r_{c}^{2} \ll 1$, a condition that is naturally satisfied when $\omega^{2} \gg G B_{0}^{2}$. The imposition of the hard-wall boundary condition leads to the following quantization condition for the energy spectrum:
\begin{equation}
\varepsilon_{\pm}=-\omega\ell\pm\sqrt{m^{2}+p_{z}^{2}+GB_{0}^{2}\ell^{2}-e\ell B_{0}+\frac{\pi^{2}}{r_{c}^{2}}\left(n+\frac{\left|\ell\right|}{2}+\frac{1}{2}\right)^{2}},
\end{equation}
where $n = 0, 1, 2, \ldots$. In this regime, the approximating $r_{\text{c}}^{2} \approx \omega^{-2}$ leads to the following energy spectrum:
\begin{equation}
    \varepsilon_{\pm}\approx-\omega\ell\pm\sqrt{m^{2}+p_{z}^{2}+GB_{0}^{2}\ell^{2}-e\ell B_{0}+\pi^{2}\omega^{2}\left(n+\frac{\left|\ell\right|}{2}+\frac{1}{2}\right)^{2}},
\end{equation}
which shows a dependence on the rotation parameter $\omega$. The term $-\omega \ell$ introduces a linear displacement associated with the angular momentum. The second term, which is under the square root, shows a more complex dependence on $\omega$, with a coupling of the rotation parameter with the quantum numbers $n$ and $\ell$.}

\begin{figure}[H]
    \centering
    \begin{subfigure}{0.49\textwidth}
        \centering
        \includegraphics[width=\linewidth]{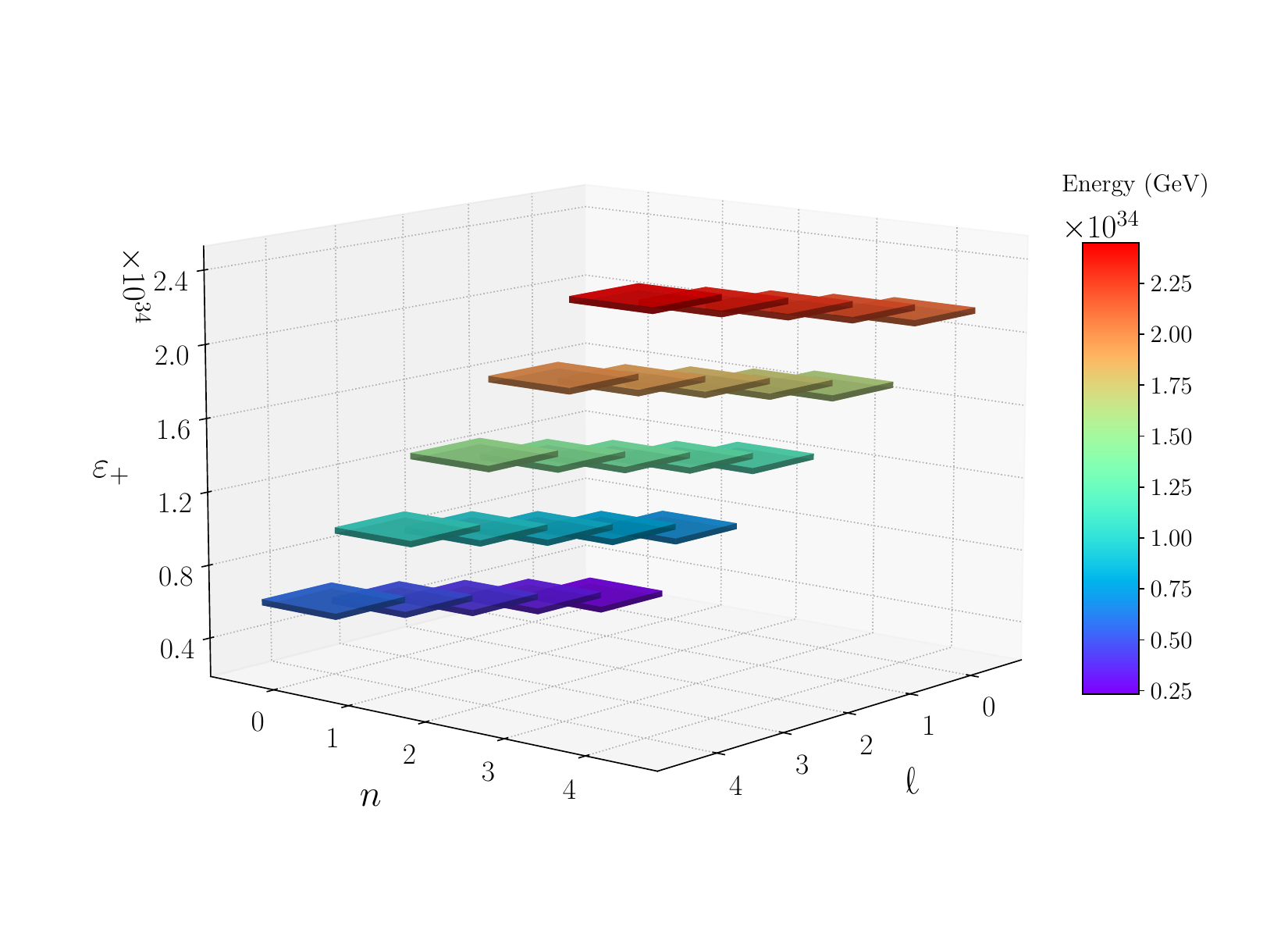}
        \caption{Positive energy levels with non-inertial effects against quantum numbers $n$ and $\ell$, for a magnetic field $B_0 = 1\,\mathrm{GeV}^2$ and $\omega=10^{-5}\omega_\mathrm{max}$, with parameters $m = m_\pi$, $e = q_e$, and $p_z = 0$.}
        \label{fig:CASEII-discrete3dnl-1gev2-noninertial}
    \end{subfigure}
    \hfill
    \begin{subfigure}{0.49\textwidth}
        \centering
        \includegraphics[width=\linewidth]{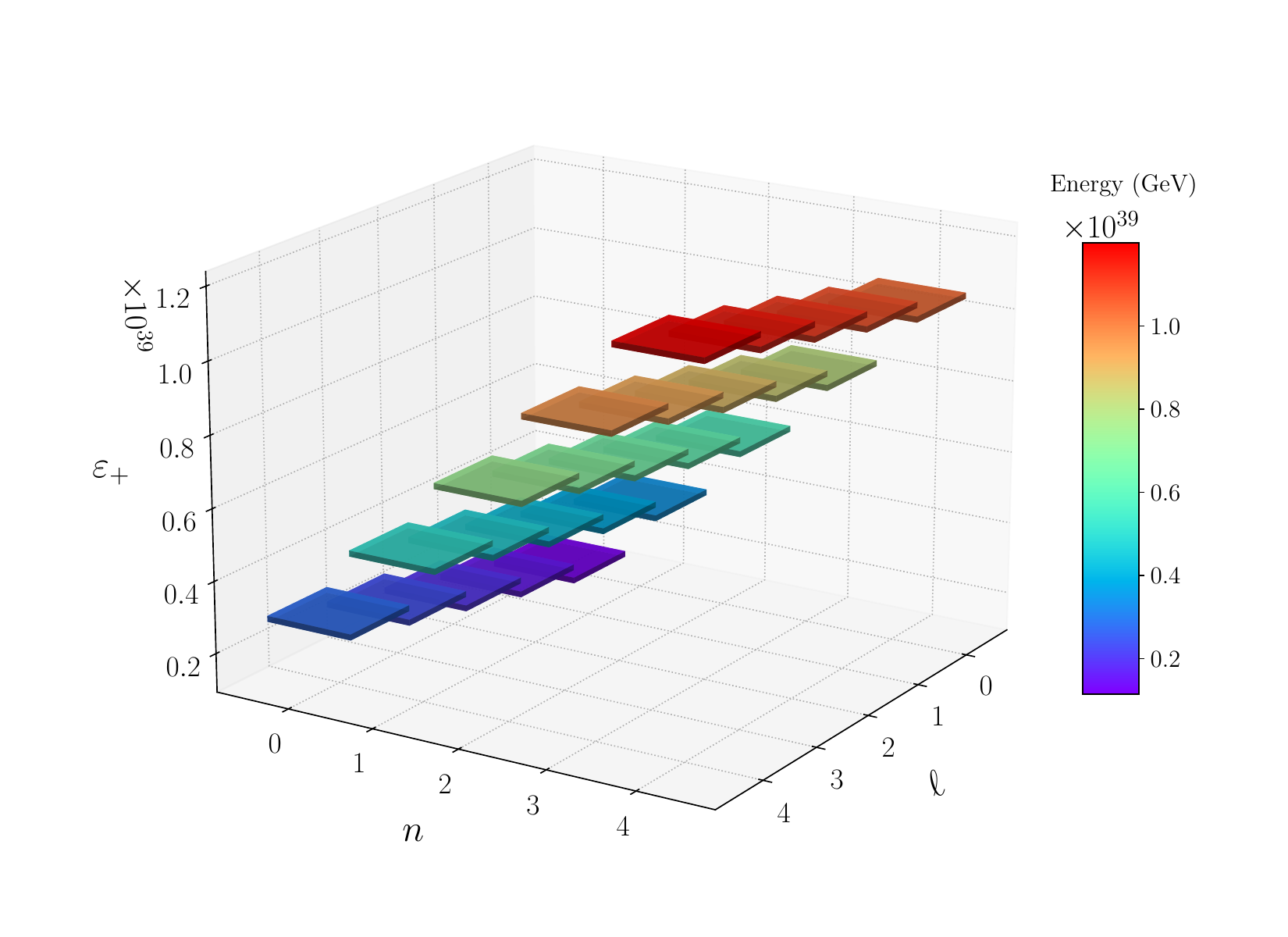}
        \caption{Positive energy levels with non-inertial effects against quantum numbers $n$ and $\ell$, for a magnetic field $B_0 = 0.1m_P^2$ and $\omega=0.5\omega_\mathrm{max}$, with parameters $m = m_\pi$, $e = q_e$, and $p_z = 0$.}
        \label{fig:CASEII-discrete3dnl-noninertial}
    \end{subfigure}
    \caption{Comparison of positive energy levels for different magnetic field strengths with non-inertial effects in spacetime.}
    \label{fig:CASEII-combined-discrete3dnl-noninertial}
\end{figure}

\textcolor{black}{For the low limit of \( r_\mathrm{c} \), we observe a specific behavior, illustrated in \autoref{fig:CASEII-combined-discrete3dnl-noninertial}. At high angular velocity scales, the angular velocity leads to a significant positive shift for positive energy values (and a corresponding negative shift for negative energy values), which dominates over the negative shift due to the term \(-\omega\ell\). This way, confining the particle to a region where the metric signature remains unchanged can result in an energy increase caused by non-inertial effects.}

\section{Conclusions}\label{Conclusions}
In this work, we have analyzed the dynamics of a charged scalar boson in the Melvin universe by solving the covariant Klein-Gordon equation with a minimal coupling. Our investigation considered both inertial and non-inertial reference frames, allowing us to explore the effects of strong magnetic fields and rotation-induced geometries, for the first time in the literature, on the quantum scalar wave equation.

By introducing a rotating frame of reference, the original Melvin spacetime was modified, leading to the emergence of a critical radius $r_c$ that bounds the physical region of the radial coordinate. Thus, in the case of the solution that takes into account the rotation of the reference frame, we observe that the particle speed could exceed the speed of light, which is not allowed. Thus, we impose that the wave function vanishes at $r_c$. The presence of this radius imposes a natural boundary condition and modifies the behavior of the wave equation, particularly. Furthermore, we obtained exact solutions for the wave equation under suitable approximations and derived the corresponding energy spectra.

The energy levels were shown to be significantly affected by the magnetic field strength, the gravitational coupling, and the non-inertial parameters. In particular, rotational effects introduced a coupling between the angular momentum quantum number and the angular velocity, shifting the energy spectrum and, in some regimes, leading to a sign inversion of the eigenvalues. These effects become especially relevant when analyzing scenarios with strong magnetic fields, such as those found in ultra-relativistic heavy-ion collisions.

The main results can be summarized as follows:

\begin{itemize}
    \item We revisited the Melvin solution of the Einstein-Maxwell equations and introduced a non-inertial (rotating) frame, leading to a modified spacetime geometry with a critical radius $r_c$.
    \item We derived analytical solutions for the Klein-Gordon equation in both inertial and rotating frames, using an appropriate ansatz and Whittaker functions.
    \item The energy spectrum shows explicit dependence on the magnetic field, gravitational corrections (via $G$), and rotational effects (via $\omega \ell$), leading to modified Landau-like levels.
    \item In the inertial system, gravitational terms introduce corrections to the Landau levels, particularly for high magnetic field intensities.
    \item In the rotating case, the energy levels are shifted by a term proportional to $\omega \ell$, with additional contributions from the modified geometry and gauge potential.
    \item Two regimes were analyzed for the non-inertial case: one where the radial domain extends to infinity (\( \omega \to \omega_{\text{min}} \)) and another where a hard-wall boundary condition is required due to a finite critical radius ($\omega_{\text{max}} \geq \omega \gg \omega_{\text{min}}$).
    \item Numerical results show that the wave function localization and energy levels are highly sensitive to changes in magnetic field strength and angular velocity, especially near the Planck scale.
\end{itemize}

These results may serve as a starting point for future investigations into scalar and fermionic wave equations in other curved and rotating geometries, potentially including the presence of additional potentials.

\section{Acknowledgments}
LGB and JVZ acknowledge the financial support from CAPES (process numbers 88887.96\\8290/2024-00 and 88887.655373/2021-00, respectively). LCNS would like to thank FAPESC for financial support under grant 735/2024. C. C. B. Jr. would like to thank CNPq.

\bibliographystyle{ieeetr}
\bibliography{sample}

\appendix

\section{Riemann Tensor Components}\label{Riemann_Tensor_Components}

In this appendix, we present the components of the Riemann tensor associated with the general metric used in this work:
\begin{equation}
  ds^{2}=\Lambda^{2}(r)\left(-dt^{2}+dr^{2}+dz^{2}\right)+\frac{r^{2}}{\Lambda^{2}(r)}d\phi^{2}
\end{equation}
\begin{equation}
    \Lambda(r)=1+\frac{1}{4}GB_{0}^{2}r^{2}
\end{equation}
where the nonvanishing components are given by
\begin{equation}
    R_{trtr}=R_{rtrt}=R_{rzzr}=R_{zrrz}=\frac{1}{2}GB_{0}^{2}\left(1-\frac{1}{4}GB_{0}^{2}r^{2}\right)
\end{equation}
\begin{equation}
   R_{trrt}=R_{rttr}=R_{rzrz}=R_{zrzr}=-\frac{1}{2}GB_{0}^{2}\left(1-\frac{1}{4}GB_{0}^{2}r^{2}\right)
\end{equation}
\begin{equation}
    R_{t\phi t\phi}=R_{\phi t\phi t}=R_{\phi zz\phi}=R_{z\phi\phi z}=\frac{GB_{0}^{2}r^{2}\left(1-\frac{1}{4}GB_{0}^{2}r^{2}\right)}{2\Lambda^{4}(r)}
\end{equation}
\begin{equation}
   R_{t\phi\phi t}=R_{\phi tt\phi}=R_{\phi z\phi z}=R_{z\phi z\phi}=-\frac{GB_{0}^{2}r^{2}\left(1-\frac{1}{4}GB_{0}^{2}r^{2}\right)}{2\Lambda^{4}(r)}
\end{equation}
\begin{equation}
    R_{tztz}=R_{ztzt}=\frac{1}{4}G^{2}B_{0}^{4}r^{2}
\end{equation}
\begin{equation}
    R_{tzzt}=R_{zttz}=-\frac{1}{4}G^{2}B_{0}^{4}r^{2}
\end{equation}
\begin{equation}
    R_{r\phi r\phi}=R_{\phi r\phi r}=\frac{512GB_{0}^{2}r^{2}\left(1-\frac{1}{8}GB_{0}^{2}r^{2}\right)}{4^{4}\Lambda^{4}(r)}
\end{equation}
\begin{equation}
    R_{r\phi\phi r}=R_{\phi rr\phi}=-\frac{512GB_{0}^{2}r^{2}\left(1-\frac{1}{8}GB_{0}^{2}r^{2}\right)}{4^{4}\Lambda^{4}(r)}
\end{equation}

Using these components, one can compute the Kretschmann scalar, which provides a measure of the spacetime curvature.

\end{document}